%% file: main_tvcg.tex
\documentclass[journal]{IEEEtran}
\usepackage{amsmath,amsfonts}
\usepackage{algorithmic}
\usepackage{algorithm}
\usepackage{array}
\usepackage[caption=false,font=normalsize,labelfont=sf,textfont=sf]{subfig}
\usepackage{textcomp}
\usepackage{stfloats}
\usepackage{url}
\usepackage{verbatim}
\usepackage{graphicx}
\usepackage{cite}
\usepackage{orcidlink}
\renewcommand{\thesection}{\arabic{section}}
\renewcommand{\thesectiondis}{\arabic{section}}
\renewcommand{\thesubsection}{\arabic{section}.\arabic{subsection}}
\renewcommand{\thesubsectiondis}{\arabic{section}.\arabic{subsection}}
\renewcommand{\thesubsubsection}{\arabic{section}.\arabic{subsection}.\arabic{subsubsection}}
\renewcommand{\thesubsubsectiondis}{\arabic{section}.\arabic{subsection}.\arabic{subsubsection}}
\usepackage[nameinlink,capitalise]{cleveref}
\usepackage[all]{hypcap} 
\crefname{figure}{Fig.}{Figs.}
\crefname{table}{Tab.}{Tabs.}
\Crefname{table}{Tbl.}{Tbls.}
\crefname{section}{Sec.}{Secs.}
\Crefname{section}{Section}{Sections}
\usepackage{enumitem}
\setlist[enumerate]{itemsep=1pt,topsep=3pt}
\setlist[itemize]{itemsep=1pt,topsep=3pt}
\setlist[description]{itemsep=1pt,topsep=3pt}

\newcommand{\subcref}[2]{\hyperref[#1]{\cref*{#1}#2}}

\newcommand{\name}{{\textit{Prismatic}}}

\usepackage{xspace,xpunctuate}
\newcommand{\ie}{\textit{i.e.,}\xspace}
\newcommand{\eg}{\textit{e.g.,}\xspace}
\newcommand{\etal}{\textit{et~al\xperiod}\xspace}

\definecolor{indicator}{RGB}{177, 47, 31}
\definecolor{explanation}{RGB}{60, 103, 143}
\definecolor{knowledge}{RGB}{91, 157, 59}
\definecolor{salmon}{rgb}{1, 0.55, 0.42}

\newcommand{\rc}[1]{\textcolor{black}{#1}}
\newcommand{\rw}[1]{\textcolor{black}{#1}}

\begin{document}
\title{Prismatic: Interactive Multi-View Cluster Analysis of Concept Stocks}

\author{
    Wong Kam-Kwai\,\orcidlink{0000-0002-2813-1972},
    Yan Luo,
    Xuanwu Yue,
    Wei Chen,
    and Huamin~Qu%
\IEEEcompsocitemizethanks{
\IEEEcompsocthanksitem W. Kam-Kwai, Y. Luo, and H. Qu are with HKUST. Email: \{kkwongar, cseluoyan, huamin\}@cse.ust.hk. H. Qu is the corresponding author.%
\IEEEcompsocthanksitem X. Yue is with Sinovation Ventures. Email: xuanwu.yue@connect.ust.hk.%
\IEEEcompsocthanksitem W. Chen are with State Key Lab of CAD\&CG and Laboratory of Art and Archaeology Image, Zhejiang University. Email: chenvis@zju.edu.cn.}%
\thanks{Manuscript received April 19, 2021; revised August 16, 2021.}
}

\markboth{Journal of \LaTeX\ Class Files,~Vol.~14, No.~8, August~2021}%
{Shell \MakeLowercase{\textit{et al.}}: A Sample Article Using IEEEtran.cls for IEEE Journals}


\maketitle

\input{0-abstract}

\input{1-introduction}
\input{2-related-works}
\input{3-background}
\input{4-system}
\input{5-design}
\input{6-evaluation}
\input{7-discussion}
\input{8-conclusion}

\section*{Acknowledgments}
We are grateful to Richard Chen, Anthony Leung, and Yang Lv for their helpful feedback. We also thank anonymous reviewers for their constructive comments. This research was supported in part by HK RGC GRF 16210321. Wei Chen is supported by the National Natural Science Foundation of China (No. 62132017), the Zhejiang Provincial Natural Science Foundation of China (No. LD24F020011), the Key R\&D ``Pioneer'' Tackling Plan Program in Zhejiang Province (No. 2023C01119).

\bibliographystyle{src/abbrv-doi-hyperref}
\bibliography{main}

\end{document}

%% file: 0-abstract.tex
\begin{abstract}
Financial cluster analysis allows investors to discover investment alternatives and avoid undertaking excessive risks. 
However, this analytical task faces substantial challenges arising from many pairwise comparisons, the dynamic correlations across time spans, and the ambiguity in deriving implications from business relational knowledge.
We propose {\name}, a visual analytics system that integrates quantitative analysis of historical performance and qualitative analysis of business relational knowledge to cluster correlated businesses interactively.
{\name} features three clustering processes: dynamic cluster generation, knowledge-based cluster exploration, and correlation-based cluster validation.
Utilizing a multi-view clustering approach, it enriches data-driven clusters with knowledge-driven similarity, providing a nuanced understanding of business correlations.
Through well-coordinated visual views, {\name} facilitates a comprehensive interpretation of intertwined quantitative and qualitative features, demonstrating its usefulness and effectiveness via case studies on formulating concept stocks and extensive interviews with domain experts.
\end{abstract}

\begin{IEEEkeywords}
Financial data, Interactive clustering, Multi-view clustering, Multi-layer network.
\end{IEEEkeywords}

%% file: 1-introduction.tex
\section{Introduction}
\label{sec:introduction}
\IEEEPARstart{C}{luster} analysis has been extensively used in analyzing financial data to understand business characteristics, portfolio optimization, and risk management~\cite{task_marti_2021}.
Economists have developed hierarchical clusters (\eg economic sectors such as energy and real estate) qualitatively to describe the affinity of different business entities based on their market coverage and product specialization~\cite{introduction_gupta_1972}. 
To address rapid market changes, professional traders have introduced \textit{concept stocks}~\cite{background_cluster_gwilym_2016}, hereafter ``concepts,'' to symbolize companies with shared business operations or similar business models in the short term.
Entities within the cluster are influenced by similar sets of economic factors that induce business-specific risks and opportunities.
Therefore, they often exhibit correlated financial performance.
In light of this, exploring the clusters' dynamics is crucial to traders, fund managers, and financial analysts to discover investment alternatives and mitigate excessive risks from investing in the same cluster~\cite{introduction_tola_2008}. 

Quantitative analysis has been applied to recreate these hierarchical clusters (\eg Sector-Industry-Company in \cref{fig:business-taxonomy}) based on stock price correlations~\cite{introduction_mantegna_1999, introduction_onnela_2003}.
Yet, investors rarely rely solely on historical stock prices when formulating investment plans. 
Instead, they adopt a holistic market view that incorporates qualitative insights about the intrinsic nature of business operations.
\rw{For instance, while stock prices provide valuable information, integrating insights from business relations (\eg the effects of market events~\cite{motivation_deng_2019} and inter-company linkages~\cite{motivation_feng_2019}) can reveal underlying dynamics that pure quantitative analysis may overlook.}
\rc{In this work, we define \textit{business relational knowledge} as the insight derived from explicit relationships among companies. This includes factors such as shared geographical regions which often indicate exposure to similar regulatory environments, strategic partnerships that suggest collaborative market behavior~\cite{related_network_basole_2013}, and supply chain connections that reveal operational dependencies~\cite{motivation_arleo_2023}. 
These qualitative factors complement quantitative data, providing a richer context for interpreting financial clusters.}

However, incorporating business relational knowledge into quantitative analysis poses substantial challenges.
First, the sheer amount of assets in the market exponentially increases the computations needed for analyzing interfirm correlations, thereby demanding considerable refinement in exploratory analysis.
Second, financial correlations are dynamic and volatile. 
During severe economic downturns, assets may exhibit synchronized stock price movements accompanied by heightened uncertainty~\cite{introduction_fenn_2011}, undermining the purpose of cluster analysis.
Third, owing to the growing complexity of business activities, the impacts of the heterogeneous business relationships and unexpected market events rely on analysts' qualitative interpretations, often lacking objective validation.

We propose {\name}, a visual analytics system that integrates quantitative analysis of historical performance and qualitative analysis of business relational knowledge to cluster correlated businesses interactively.
{\name} enables interactive cluster analysis with three key analytical processes: 
1) cluster generation by holistically overviewing the dynamic data-driven clusters, 
2) cluster exploration by contextualizing the clusters with business relational knowledge, and 
3) cluster validation by analyzing temporal correlation patterns at different time scales and time horizons.
Qualitative analysis within {\name} relies on business relational knowledge formulated in a multi-layer network. 
We employed a multi-view clustering method to consolidate the multiple facets and augment correlation-based clusters with businesses having high knowledge-driven similarity.
We designed a suite of well-coordinated views to integrate the results and facilitate a thorough interpretation of the interplay between qualitative and quantitative features. 
We evaluated the usefulness and effectiveness of {\name} through two case studies and interviews with domain experts. 
This paper has the following major contributions:
\begin{itemize}
\item A visual analytics approach that clusters correlated assets and \rw{relational knowledge}, bridging the gap between qualitative and quantitative analyses in the finance domain.
\item Coherent visualizations that support the entire workflow of cluster generation, exploration, and validation.
\item Case studies that demonstrate its effectiveness in exploring useful financial clusters based on users' needs.
\end{itemize}

%% file: 2-related-works.tex
\section{Related works}
This manuscript is related to interactive clustering analysis for strongly correlated securities, either from the temporal data-driven or the relational knowledge-driven perspectives.

\subsection{Interactive visual cluster analysis}
Interactive clustering~\cite{related_clustering_bae_2020} is characterized by supporting clustering algorithms with human-in-the-loop to enhance clustering quality in terms of meaningfulness, agreeableness, and interpretability. 
Users interact mainly with two components: \textit{parameter} and \textit{result interaction}.
Parameter interaction focuses on refining model and parameter selections for better initialization.
For example, \textit{Clustrophile 2}~\cite{related_clustering_cavallo_2019} and \textit{Clustervision}~\cite{related_clustering_kown_2018} visualize parameter search spaces to adjust different clustering models and parameter settings. 
However, their complexity requires expertise in underlying algorithms, potentially limiting their broader adoption in other domains.

Result interaction leverages users' domain knowledge to refine the automated clustering outputs. 
For \textit{StratomeX}~\cite{related_clustering_kern_2017} and \textit{Geono-Cluster}~\cite{related_clustering_das_2021}, analysts can steer the clustering process with interim results and performing early refinements. 
It has shown promise in clustering time series data for temporal relationships~\cite{related_clustering_turkay_2011, related_clustering_sacha_2018} and multidimensional attributes for hierarchical relationships~\cite{related_clustering_cao_2010, related_clustering_cao_2011, related_clustering_xia_2023}.
Nevertheless, these approaches have been challenged by diluting the purity of the data-driven process with users' bias~\cite{related_clustering_pister_2021}.

\textit{Knowledge assistance}~\cite{related_knowledge_nam_2009, related_knowledge_federico_2017} is one answer of high potential to the criticism by making tacit domain knowledge explicit and transferable, thereby enhancing clustering result reproducibility.
Mixed-initiative workflows like \textit{PK-clustering}~\cite{related_clustering_pister_2021} and \textit{CohortVA}~\cite{zhang2023cohortva} cross-validate clusters with multi-faceted prior knowledge.
\textit{IRVINE}~\cite{related_knowledge_eirich_2022} features tacit knowledge annotation and collaborative data labeling with multiple users.
These few works shape a novel approach to enhancing interactive clustering results.
However, they primarily address hierarchical or dynamic relationships, not both, which are equally important in the financial domain.
Our work combines the best of both worlds, exploring interactive clustering with data-driven and knowledge-assisted configurations.

\subsection{Financial stock data visualization}
Visualizing financial data has supported numerous business applications, including investment management~\cite{related_stock_app_yue_2020, chen2024fmlens}, tax administration~\cite{related_network_lin_2021, yuan2023taxscheduler, wong2023anchorage}, and social media analysis~\cite{related_stock_app_hullman_2013, related_stock_app_shi_2019}.
The surveys~\cite{related_stock_survey_flood_2016, related_stock_survey_ko_2016} have summarized the research horizons on financial visualization about their positive impact in facilitating efficient financial decision-making.
This work focuses on the stock market, where time series data describe the temporal evolution of business entities' valuation.

Visualizing from a \textit{market perspective} is technically challenging because of the massive number of stocks in a market.
Treepmap~\cite{related_stock_market_wattenberg_1999} and ring-based layout~\cite{related_stock_market_lei_2010} were developed to outline the whole market leveraging the concept of industry and market index.
Yet, these visualizations rely on predefined industry hierarchies and overlook their dynamic nature in the complex business world.
To mitigate the issue, Ziegler~\etal~\cite{related_stock_market_ziegler_2010} explored small multiples of heatmaps to juxtapose financial time series across different market sectors. 
3D heatmaps~\cite{related_stock_market_joseph_2013} and dendrograms~\cite{related_stock_market_rea_2014} are common designs for visualizing correlated stocks.
Nevertheless, they consider only price or capitalization correlations susceptible to short-term fluctuations. 
Their visual designs fail to encode auxiliary information to evaluate companies comprehensively.

Another thread of research focuses on the individual \textit{stock perspectives}.
Keim~\etal~\cite{related_stock_pairwise_keim_2006} proposed the growth matrix technique to compare multiple assets over all possible sub-intervals.
Alsakran~\etal~\cite{related_stock_pairwise_alsakran_2010} developed a tile-based parallel coordinate plot to superpose time series with the most frequency and remove outliers.
These methods show multi-scale temporal trends and offer pairwise comparisons for financial products.
Thermal~\cite{related_stock_pairwise_stitz_2016} and swarm metaphor~\cite{related_stock_pairwise_simon_2018} supported the interactive exploration of stock performance with heatmaps, stream graphs, and scatterplots. 
While these works emphasize the dynamic evolution of stocks with animated displays, they cannot support human-in-the-loop reasoning about correlation patterns. 
Our work expands upon these ideas by showcasing temporal development from market, cluster, and individual stock perspectives.
The visual components of {\name} are designed to find clusters of correlated assets across varying time frames and multiple time horizons.

\subsection{Visual analytics for business networks}
Mining closely connected entities from business networks fosters a better understanding of business nature and activities.
\textit{TaxThemis}~\cite{related_network_lin_2021} investigates suspicious tax evasion groups conducting related party transactions with ownership networks and transaction records. 
Similarly, \textit{NEVA}~\cite{related_network_leite_2020} and \textit{iConViz}~\cite{related_network_niu_2020} detect financial fraud based on banks' client profiles.
These works are helpful in fraud detection using the transactional data provided by authorities.
Still, they have limited transferability to analyzing the stock market, where business relationships are more complex than transactions~\cite{related_network_basole_2017} and often undisclosed~\cite{motivation_arleo_2023}. 
Using these systems directly might suffer partial understanding and insufficient evidence.

To reveal the hidden relationships in business networks, \textit{Ecoxight}~\cite{related_network_basole_2018} and \textit{dotlink360}~\cite{related_network_basole_2013} adopt contract data from commercial and public databases to examine business ecosystems. 
Decentralized temporal transaction data has been explored to demonstrate the cryptocurrency exchanges' development and collaborations~\cite{related_network_yue_2019, chen2024ethereum}.
Similar to these systems, we explore and refine clusters of strongly correlated securities by combining publicly available heterogeneous data.
However, {\name} differs from previous works in which the relational data is formulated into a multi-layer network.
We leverage multi-view clustering to aggregate the diverse complementary relationships and generate knowledge-driven clusters.

%% file: 3-background.tex
\section{Background}

This section provides background information about cluster analysis in the finance industry.
It also illustrates the data and task abstraction, following the design study guidelines~\cite{task_munzner_2009}.

\subsection{Cluster analysis for concept stocks}
\begin{figure}[tb]
    \setlength{\belowcaptionskip}{-0.3cm}
	\centering
	\includegraphics[width=\linewidth]{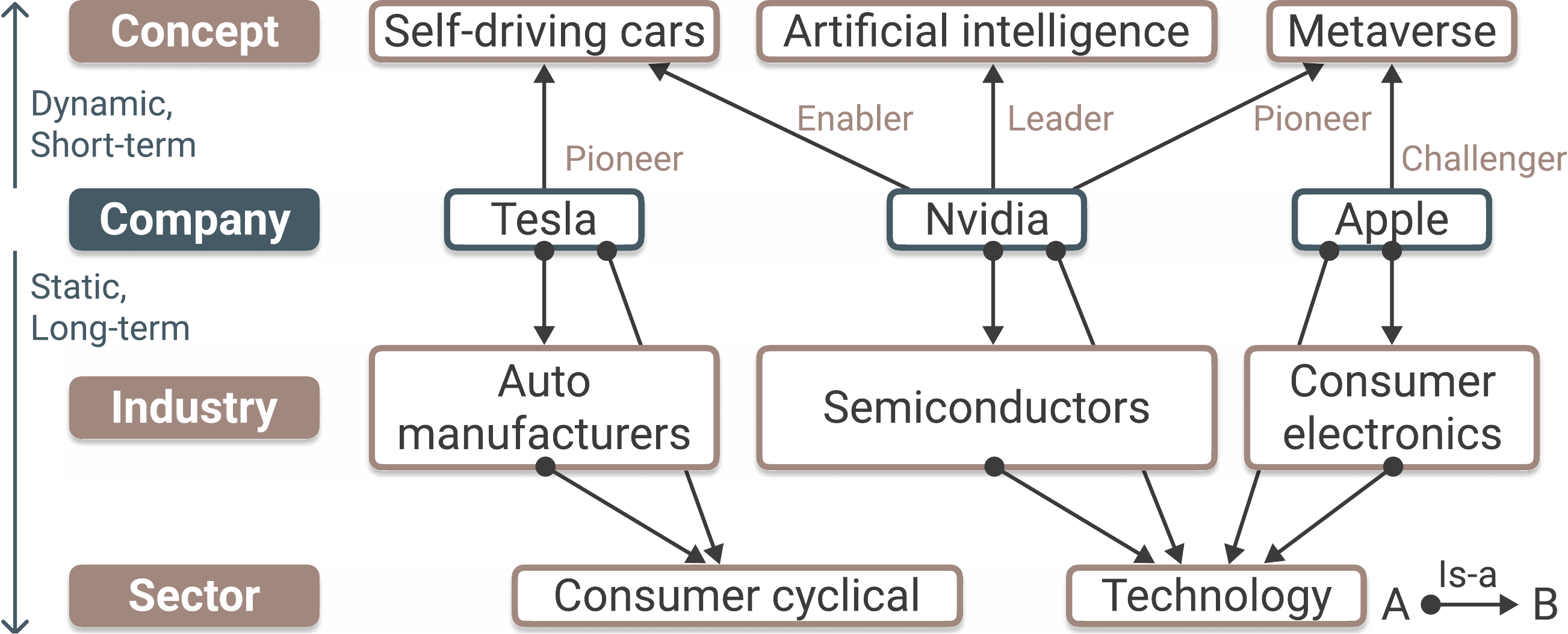}
    \caption{A simplified example illustrating different financial clusters. Sectors and industries are related hierarchically, while concept stocks can be constructed arbitrarily to label specific \rw{business relational knowledge.}}
    \label{fig:business-taxonomy}
\end{figure}
Traditional economic sector classification forms the basis of financial clusters utilized in equity research~\cite{introduction_gupta_1972}, with further subdivision into industries to capture more nuanced business characteristics.
This established business taxonomy, characterized by a static hierarchical structure, can remain unchanged for decades~\cite{task_marti_2021}.
However, it falls short in classifying fast-evolving businesses, especially in developing economies where business natures may change swiftly.
In light of this, \textit{concept stocks}, as a more dynamic and adaptable approach, have been introduced to group stocks based on their association with emerging, often disruptive, technological trends, innovative business models, and specific thematic developments, as illustrated in \cref{fig:business-taxonomy}.
For example, Nvidia is a core enabler for the self-driving cars concept as it provides computation infrastructure for many practitioners.
Yet, it may be exempted from the concept if all practitioners decide to build their own computing units.
Over time, such emerging concepts may mature, forming new industries in traditional economic sectors, such as the internet and electrical automobile industries.

The generation of these concepts is neither systematic nor hierarchical, often resembling phenomena or hypotheses with anecdotal definitions.
These concepts act as proxies reflecting retail investors' speculative interest and attention~\cite{background_cluster_gwilym_2016}, fostering strong correlations in stock prices and trading volumes.
For instance, the Chinese `pork' concept emerged from the observation that pork prices have a predominant influence on stock prices with statistical significance. 
Companies within the pork concept or along its supply chain are heavily influenced by pork prices, independent of investor sentiment or macroeconomic factors~\cite{background_cluster_liu_2020}.
Therefore, similar to traditional clusters, concepts serve to hedge risks.
Investing in multiple companies within the pork concept could entail excessive risk when pork prices drop.
This fresh perspective of viewing stocks through the lens of concept offers a dynamic approach to understanding market correlations and evolving business landscapes.
It aids in capturing potentially significant market trends that traditional sector-based cluster analysis might otherwise overlook.


\subsection{Task analysis}
Integrating quantitative approaches for financial correlation analysis~\cite{related_stock_pairwise_simon_2018} with knowledge-assisted clustering remains largely unexplored.
We first surveyed through literature about \textit{quantitative analysis}, which is the study of financial products using mathematical and statistical methods, to obtain the technical background and lay down the analytical framework for financial correlation analysis.
We then interviewed three financial analysts from different investment research firms to understand the existing analysis workflows.
With at least five years of experience in \textit{fundamental analysis}, these analysts provided insights into the qualitative asset valuation approach focused on financial statements and market factors.
Under the influence of visual comparison considerations~\cite{task_gleicher_2018} and the interactive clustering paradigm~\cite{related_clustering_bae_2020}, we summarized the design requirements for developing strongly correlated clusters.
Three stages of design requirements are identified as follows.


\textbf{Cluster Generation (CG)} describes the model usage and proposes new clusters.
A meaningful cluster is partitioned from the financial market by clustering methods and is the hypothesis for subsequent analysis.
\begin{enumerate}[label=\textbf{CG{\arabic*}}]
    \item \textbf{Identify closely correlated stocks as cluster candidates.} 
    The number of securities in the financial market easily exceeds the analysts' capability to process at once, requiring aid in finding an appropriate-sized cluster.
    It is a non-trivial task because, according to all experts, there could be thousands of stocks simply moving along with the market index.
    These `market sentiments' can bring noise to the impacts of stock- or industry-specific events~\cite{related_stock_pairwise_simon_2018}.
    We should provide a multi-level measurement to avoid disturbance and locate strongly correlated stocks.
    \item \textbf{Summarize the correlation dynamics of the clusters.} 
    Under different economic cycles, the market and securities show different correlation patterns, given their economic positions.
    In volatile markets like a financial crisis, most stocks show almost identical movements and considerable uncertain behaviors.
    It would significantly distort the interpretation of the underlying interrelationships~\cite{system_correlation_aste_2010}.
    We should provide an overview of the market and cluster correlation so that users can connect the dynamic relationships across time frames and choose the investigation period of interest~\cite{related_stock_pairwise_simon_2018}.
\end{enumerate}

\textbf{Cluster Exploration (CE)} augments the correlation-based clusters with business relational knowledge.
It should expand the selected cluster with qualitatively similar businesses.

\begin{enumerate}[label=\textbf{CE{\arabic*}}]
    \item \textbf{Explore business relationships outside the cluster.} 
    Due to their dynamic nature, correlation-based clusters are over-sensitive to short-term fluctuations and sub-optimal time frame configuration.
    Alternative data (\eg supply chains, policies, and critical personnel data) can provide a relatively static grounding to understand the correlations~\cite{task_marti_2021} and prompt for associating prior knowledge~\cite{related_clustering_pister_2021}.
    They can also be used to model business proximity~\cite{related_network_basole_2013} and cross-validate common relationships~\cite{zhang2023cohortva}.
    We should provide an interactive subset selection of business relationships for knowledge-assisted clustering.
    \item \textbf{Summarize data-driven and knowledge-driven perspectives.}
    Analyzing complex company relationships requires significant effort due to the vast number of possible connections.
    The lack of standardized measurement for various knowledge types can lead to heavy reliance on analysts' subjective interpretation.
    Knowledge elicitation~\cite{related_knowledge_eirich_2022} can offset the subjectivity and reduce the burden on users' cognitive loads.
    We should provide a visual summary of the data-driven correlation and knowledge-driven business relationships such that the two perspectives can complement each other for better valuation.
\end{enumerate}

\textbf{Cluster Validation (CV)} focuses on the confirmation of the explored hypothesis. This process should identify unrelated or dissimilar businesses and shrink the cluster.
\begin{enumerate}[label=\textbf{CV{\arabic*}}]
    \item \textbf{Summarize the intra-cluster correlations.}
    The selected cluster should attain an acceptable level of correlation among its members.
    Ill-defined model configuration could cause unsatisfactory results.
    We should provide a visual summary for judging whether to interact with the model's parameters or proceed with amending the result~\cite{related_clustering_bae_2020}.
    It also reminds users about the completion progress.
    \item \textbf{Identify the weakest relationship for validation.}
    From the experts' point of view, a sound financial cluster should contain all highly correlated stocks.
    A greedy approach removes the lowest correlated stock or one with the weakest business relationship from the cluster. 
    The weakest relationship is often evaluated based on stock prices.
    In reality, investors would also examine other financial variables, such as trading volume~\cite{task_podobnik_2009}, for more information.
    We should provide auxiliary information to contextualize the weakest relationships, which could also be used as tie-breaking rules for similar price correlations.
    \item \textbf{Dissect the pairwise comparison of correlation in multi-scale.}
    The dynamic nature of correlation complicates pairwise comparisons between stocks; thus, multi-scale time frames should be taken into consideration. 
    The pairwise comparison might occur between entities that users are not familiar with, thereby requiring a common reference point for comparisons.
    We should provide a multi-hierarchical and multi-scale temporal comparison to facilitate the validation decision.
\end{enumerate}

%% file: 4-system.tex
\section{The Framework of Prismatic}
\begin{figure*}[tb]
    \setlength{\belowcaptionskip}{-0.3cm}
    \centering
    \includegraphics[width=\linewidth]{figures/framework-compressed-v2.pdf}
    \caption{
    Overview of the {\name} framework. The example follows from \cref{fig:business-taxonomy}. The three core modules of {\name} enhance user exploration of concept stocks through a visual interface.
    (A) The data-driven perspective constructs a financial correlation network from financial time series to support multi-scale exploration of correlations. This involves (A1) checking for monotonic price co-movements, (A2) filtering out weak correlations to focus on significant relationships, and (A3) identifying key companies using betweenness centrality.
    (B) The knowledge-driven perspective constructs a multi-layer network from \rw{business relational knowledge} and produces multi-view clusters. This network contains factual information about the business relationships, while the clusters assess business proximity through various business domain perspectives.
    (C) {\name} interface offers an integrated workflow for interactive visual cluster analysis with cluster generation, summary, exploration, and validation, facilitating a seamless integration of data-driven and knowledge-driven insights.}
    \label{fig:framework}
\end{figure*}
\rw{The framework of {\name} (\cref{fig:framework}) consists of three modules.
To generate cluster candidates, we first compute yearly correlation networks to analyze cluster composition (\textbf{CG2}, \subcref{fig:framework}{C1}).
These networks are piped into community detection for correlated stocks (\textbf{CG1}, \subcref{fig:framework}{A}) and multi-view clustering to retrieve related businesses (\textbf{CE1}, \subcref{fig:framework}{B}) to form an initial financial cluster template.
The knowledge-driven perspective provides ego-centric exploration, allowing users to refine clusters by including related stocks (\textbf{CE1}, \subcref{fig:framework}{C3}).
The cluster summary highlights intra-cluster correlations and shared knowledge (\textbf{CE2}, \textbf{CV1}, \subcref{fig:framework}{C2}), helping users identify weakly correlated stocks for cluster validation (\textbf{CV2}).
The Prism time series view supports pairwise comparisons and data-driven validation of correlated stocks (\textbf{CV3}, \subcref{fig:framework}{C4}).}

\subsection{Financial correlation network}
We first constructed the correlation network to highlight price comovements with companies as the nodes and their Spearman and Pearson correlations from both stock prices and trading volumes as the edges' attributes (\subcref{fig:framework}{A}).
Pearson correlation primarily measures linear dependencies and is prone to outliers~\cite{system_data_guo_2018}, while Spearman correlation measures monotonic relationships and is less sensitive to outliers as their values are bounded by rank, especially when the correlation is relatively strong~\cite{system_correlation_eisinga_2013}.
In cluster generation, the Spearman correlation checks for monotonic relationships (\subcref{fig:framework}{A1}) before using Pearson correlation to reduce potentially weakly correlated companies (\subcref{fig:framework}{A2}).

To identify strongly correlated communities, we pruned less correlated nodes by user-specified thresholds. 
While we focused on correlation coefficients, other criteria, such as statistical significance, can also guide the pruning process, allowing users to analyze an appropriate-sized, relatively dense cluster.
The computing time can also benefit from the reduced number of components, a quadratic growth factor for comparisons.
Detecting groups using simple similarity metrics could generate a large sub-network because a few ``super-connectors'' provide linkage between subgroups~\cite{related_network_lin_2021}.
We leveraged the betweenness centrality~\cite{system_betweenness_freeman_1977} to find such super-connectors and measure the node importance in the cluster (\subcref{fig:framework}{A3}).

\subsection{\rw{Business relational knowledge mining}}
\label{sec:knowledge}
\rw{We acquired stock data and business relational information from Tushare~\footnote{\url{https://tushare.pro/}}, a community-maintained database for the Chinese financial market~\cite{motivation_chen_2018}.
The dataset covers Chinese A-share listed firms in the Main Board and the Small and Medium Enterprises (SME) Board of the Shanghai and Shenzhen stock exchanges between 2010 and 2020, resulting in 2875 stocks.
This period contains different phases of economic cycles, such as stagnation in 2014, large volatility in 2015, and recovery in 2016~\cite{system_data_guo_2018}.
In addition to collecting split-adjusted daily closing prices and trading volumes for quantitative analysis, we obtained profile information such as company locations, business activities, management details, and top investor lists.}

To model and capture business relational knowledge, we identified three critical dimensions from our dataset: human, location, and business factors (\subcref{fig:framework}{B1}).
These dimensions were fused into a multi-layer network to provide a structured representation across data sources~\cite{system_network_mcgee_2019}. 
In our network, each of the 2875 nodes represents a listed company, and approximately 3.1 million edges encode the relationships across the three dimensions. 
Each dimension is characterized by primary and secondary attributes, and companies are linked based on shared features within each dimension.

Specifically, the location layer utilizes province and city information to reflect the influence of regional environments and regulatory policies, thereby capturing geographical similarities that may affect company performance.
The human layer focuses on shared investors and management, which can reveal close business affiliations and strategic partnerships~\cite{motivation_chen_2018}.
To mitigate issues of name ambiguity, especially among Chinese names~\cite{system_data_shen_2017}, we employed name disambiguation techniques that consider both full names and birth years.
The business layer includes industry classifications and concepts obtained from investment research firms, offering a granular description of each company’s operational focus.
Industry classifications were sourced from Shenyin Wanguo Securities Research, and concepts from RoyalFlush Network Technology.
\subcref{fig:framework}{B} demonstrates an illustrative example in the United States market using the primary attribute of the three layers.

\subsection{Multi-view clustering for \rw{business relational knowledge}}
We applied Graph-based Multi-view Clustering (GMC)~\cite{system_cluster_wang_2020} with the multi-layer network to enhance qualitative understanding of interrelationships among listed companies (\subcref{fig:framework}{B2}).
The resulting knowledge-driven clusters can further differentiate stocks with similar correlations, potentially reducing cluster sizes.
GMC proposes shifting conventional clustering paradigm from single to multi-view learning~\cite{system_cluster_chen_2022}, promoting a holistic approach that suits the multi-view nature of our multi-layer network.
To cope with distinct biases in each view, multi-view learning adheres to two main principles: \textit{consensus} and \textit{complementarity}.
Following the \textbf{consensus principle}, GMC first learns an appropriate similarity metric for each layer, called the Similarity-Induced Graph (SIG), and then fuses these SIGs to maximize the unified graph by monitoring the agreements among different layers.
For instance, firms in the finance industry located in Shanghai could leverage distinct SIGs from both business and location layers to merge specific business insights with regional benefits.
Furthermore, adhering to the \textbf{complementary principle}, GMC back-propagates the unified graph to refine SIGs.
This iterative refinement lets complementary information from different layers spread through each layer, adjusting their weights to signify their relevance and improve clustering outcomes based on the unified graph.
The implementation was modified from an open-sourced repository~\footnote{\url{https://github.com/cshaowang/gmc}}.

%% file: 5-design.tex
\begin{figure*}[tb]
    \setlength{\belowcaptionskip}{-0.3cm}
	\centering
	\includegraphics[width=\linewidth]{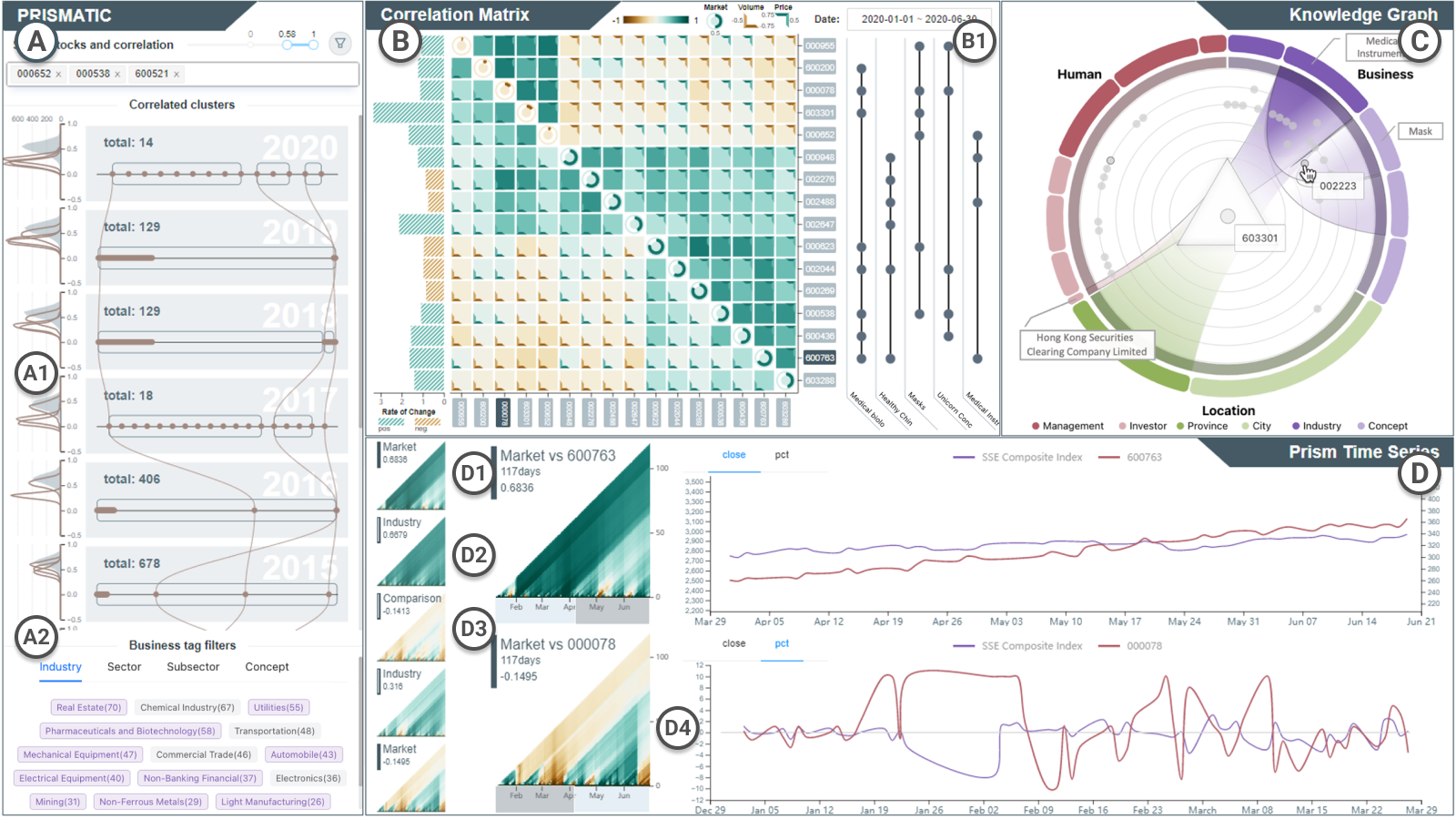}
    \caption{The visual interface of {\name}. (A) provides the model configuration and illustrates the dynamic structure of financial correlation. (B) visualizes the data-driven community and presents a highly correlated cluster after validation and exploration. (C) features the Prism view with interaction making use of the prism metaphor. (D) incorporates business relational knowledge into the system to support cluster generation.}
    \label{fig:teaser}
\end{figure*}

\section{The Visual Interface of Prismatic}
There are four main views for the visual interface of {\name} to facilitate the interactive clustering process.

\subsection{Financial correlation network view}
This view (\subcref{fig:teaser}{A}) configures the financial correlation network with yearly pre-computed correlations between all stocks.
Partitioning the correlation annually is an established practice due to its seasonal nature~\cite{design_network_loretan_2000}.
The SSE Composite Index serves as the market benchmark in this analysis.
The top control panel customizes the community detection algorithm (\textbf{CG1}), allowing users to specify ``must-have'' stocks and correlation thresholds.
The resulted correlation network is visualized in rows, each corresponding to a year and summarizing the evolving correlation structure (\textbf{CG2}, \subcref{fig:teaser}{A1}).
This vertical arrangement is referred to as ``linearized nodes on vertical axes'' for dynamic network visualizations~\cite{design_network_beck_2017}.

An overview of the correlation distribution is depicted to the left of each row. 
While the shaded area illustrates the market index's correlation distribution against all other stocks, the lines represent the selected stocks' counterparts.
The lines are superposed for comparison and offer a gross estimate of the year's correlated performance.
To the right, each row shows a rectangle representing the filtered sub-network.
A bounding box represents a correlation-based cluster, with dots inside belonging to the same cluster.
Dots are hidden if the corresponding stock is not part of the same knowledge-driven cluster of any selected stocks. 
These groups are sorted by size.
The influential stocks in the group, as determined by betweenness, are positioned on the left.
Connections between rows illustrate changes in stock positions over the years, arranged in descending order to highlight recent market trends.
At the bottom of the view, business tag filters (\subcref{fig:teaser}{A2}) based on industry classification allow further refinement based on user preferences. 
Selecting a row indicates the user's interest in exploring the cluster for that year, which is the basis for further exploration and validation.

\subsection{Correlation matrix view}
This view (\subcref{fig:teaser}{B}) presents a visual summary of the correlations among selected stocks within the cluster (\textbf{CV1}), assisting users in identifying the cluster's structures over specific time frames.
It contextualizes correlations related to price, trading volume, and the market index, facilitating the identification of group dynamics within clusters.

The view combines three distinct elements: a right-aligned bar chart for stock return to the left (\subcref{fig:teaser}{B}), a correlation matrix in the middle, and the UpSet plot~\cite{design_matrix_lex_2014} to the right (\subcref{fig:teaser}{B1}) that depicts set-based \rw{business relational knowledge} derived from the multi-layer network.
In the UpSet plot, each column represents a \rw{business relational knowledge} item, with dots indicating the presence of these items in business entities.
The left bar chart highlights stock performance, assisting in identifying the least related stocks (\textbf{CV2}).
This auxiliary information enhances distinguishability and guides users in prioritizing their focus.
Positive and negative values within this view are indicated in green and brown, respectively, deviating from traditional color schemes to enhance clarity.

The correlation matrix encodes three types of complementary information: price, volume, and market correlations.
Price correlations are positioned in the upper right triangle, while volume correlations are in the lower left.
This arrangement reflects the finance domain's interest in the interplay between trading volume and price changes due to their complementary nature~\cite{task_podobnik_2009}.
Both correlations employ color encoding to represent correlation values, while the diagonal--all ones by definition--shows each stock's correlation with the market index.
Although correlation is not transitive, we attempt to offer this information as a common reference point, contextualizing the interrelationships between different stocks and market trends.
Colors in the diagonal denote positive or negative correlations, with the correlation value encoded within the colored segment (\ie clockwise angle from zero) of the donut chart.
This design creates visually distinguishable patterns for high correlations and guides user attention effectively.

We implemented a two-phase hierarchical clustering to reveal internal structures and the inherent hierarchy among stocks.
The initial phase concentrates on stock prices to provide a clear cluster overview.
Subsequently, trading volume is considered to highlight the significance of larger companies.
This approach autonomously identifies highly correlated stock clusters, aligning with financial practices of categorizing assets into sub-classes~\cite{system_correlation_tumminello_2010}.
The matrix can be reordered through draggable labels on both axes for specific analytic purposes.

\textbf{Justification.}
The symmetry inherent in the correlation matrix is a critical element of our visualization design, resonating with the conventions familiar to economists and financial analysts.
These professionals commonly plot various coefficients within the same matrix, such as Pearson correlation above the diagonal and Spearman's rank correlation below~\cite{design_matrix_ball_1969}. 
Our design matches the convention and thus induces a limited impact on the learning curve.
Despite this familiarity, we recognize the risk of misinterpretation, which is especially when users focus on a single column or row to explore the entirety of correlations, which might overlook the distinction between price and volume correlations.

We change the diagonal cells into circular shapes to mitigate this issue and facilitate a clearer mental model. 
We also introduce an additional visual cue, the cornered border reminiscent of photo frames, that explicitly indicates price correlations by pointing towards the top right and volume the bottom left. 
This further differentiates the two regions and provides redundancy to avoid misunderstanding.
Moreover, these cornered borders serve as embedded legends superposed within the cells.
When the correlation values match the legend, their colors are also matched and blended, enabling users to estimate correlation values through direct color comparison.
This design rhymes with the design principle of eyes beat memory~\cite{related_network_yue_2019}.
The presence of numerous visible cornered borders suggests areas where the matrix might be incomplete or require further refinement (\textbf{CV2}).

\begin{figure}[tb]
    \setlength{\belowcaptionskip}{-0.3cm}
	\centering
	\includegraphics[width=\linewidth]{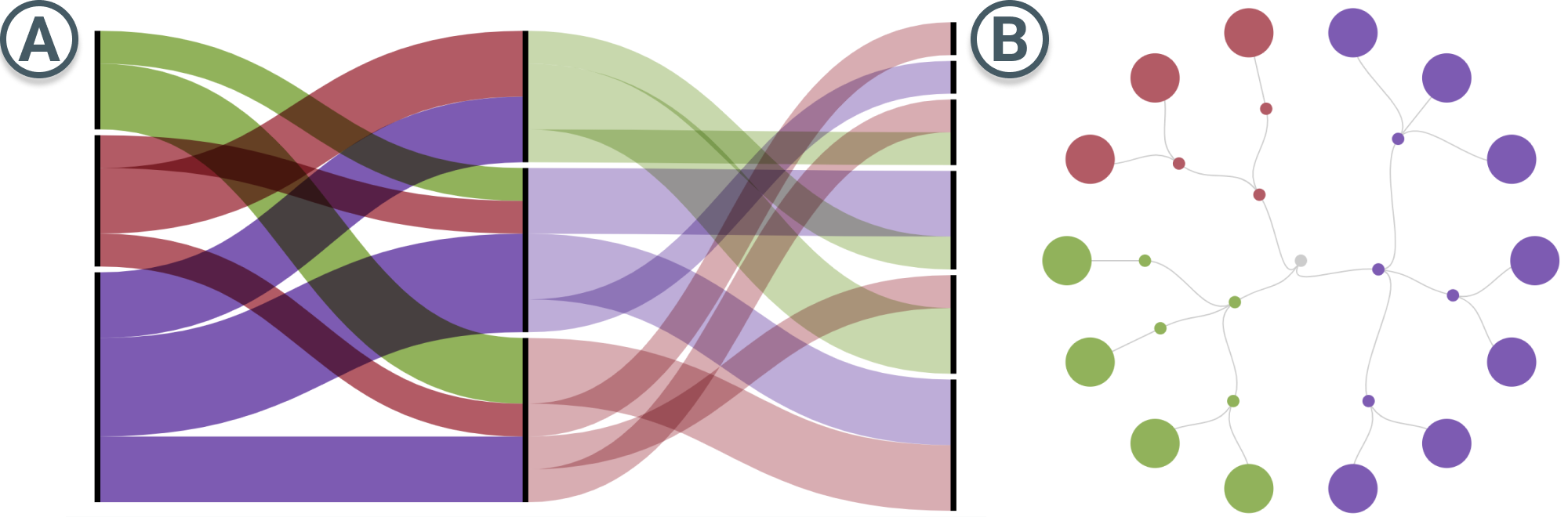}
    \caption{The alternative design for the knowledge graph view: (A) Sankey diagram and (B) circular dendrogram.}
    \label{fig:alt-design}
\end{figure}
\subsection{Knowledge graph view}
This view (\subcref{fig:teaser}{C}) activates when a stock is selected from the correlation matrix, initiating an ego-centric search in the multi-layer network for \rw{business relational knowledge} (\textbf{CE1}).
Utilizing a chord diagram variant, this view associates different knowledge items compactly~\cite{zhang2024scrolltimes}.
The circle is divided into three equal sections for three distinct knowledge layers in the multi-layer network (\cref{sec:knowledge}).
Each segment on the outer ring represents an item from a knowledge layer, with the primary attributes highlighted in darker shades.
The segment width is proportionate to the logarithm of its count and inversely to the total, highlighting rarer items by increasing their visibility. 
This design emphasizes the value of unique knowledge items, which are generally more informative, by diminishing the presence of more common ones.

Selecting a segment reveals other stocks with the same business attribute or belonging to the same cluster as determined by multi-view clustering. 
These stocks appear in the inner rings, whose arrangement indicates the number of shared knowledge items, such that the more shared relationships, the closer the node is to the center.
Users can click on a single node to highlight other shared knowledge items with the center node.
For instance, selecting a node reveals other companies sharing the same industry classification, concepts, and investors with bands passing through the center node. 
This visualization metaphorically represents the discovery process as light refracting through a prism, revealing interconnected business relationships across knowledge layers.

To contextualize the correlation matrix and summarize \rw{business relational knowledge} (\textbf{CE2}), users can double-click the knowledge item to augment the UpSet plot (\subcref{fig:teaser}{B1}) with new knowledge and then examine the business proximity between stocks in the cluster.
Likewise, double-clicking the nodes in the inner rings would add the stock to the correlation matrix, contributing to cluster exploration.

\textbf{Alternative design.} We have also considered other visualization designs for the multi-layer network. 
The Sankey diagram (\subcref{fig:alt-design}{A}) has the advantage of interactive power and the combination of the set concept.
However, it struggles with comparisons in dimensionality beyond two.
Another design is the circular dendrogram (\subcref{fig:alt-design}{B}), in which the hierarchical structure is clear, and the closeness to the center could also encode relational proximity.
Yet, the dendrogram fails to provide consistent structure when the number of knowledge increases.
Our chord diagram design maintains a balance between information clarity and structural coherence.

\begin{figure}[tb]
    \setlength{\belowcaptionskip}{-0.3cm}
	\centering
	\includegraphics[width=\linewidth]{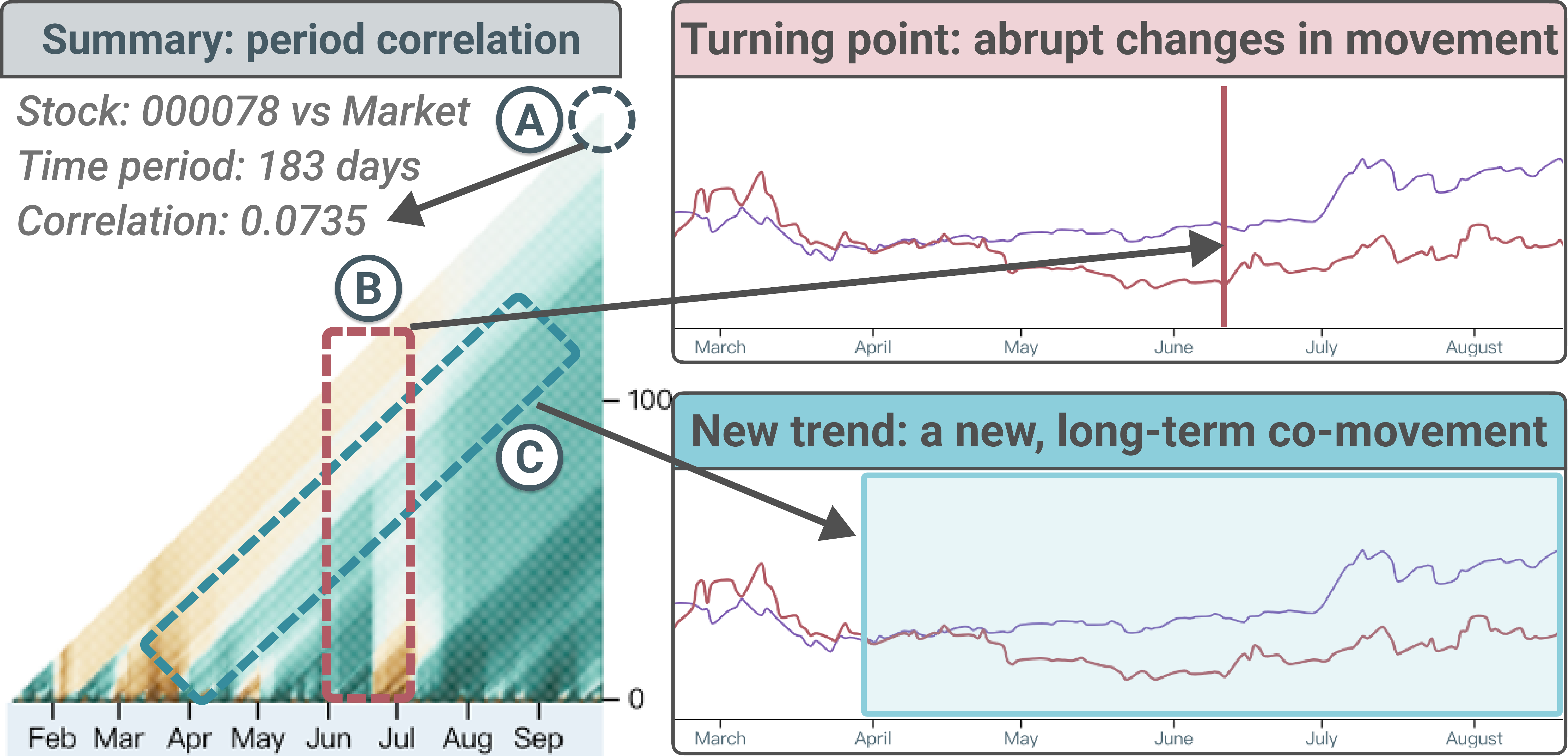}
    \caption{The Prism view design summarizes statistical aggregates across all time scales in a heatmap, with interactive features allowing users to explore and identify key trends between two time series efficiently.}
    \label{fig:design}
\end{figure}

\subsection{Prism time series view}
This view (\subcref{fig:teaser}{D}) features a pixel-based two-dimensional heatmap with a creative encoding strategy along its axes.
Each mark represents the correlation coefficient of the day and can be hovered for information.
Inspired by Keim~\etal~\cite{related_stock_pairwise_keim_2006} and Sips~\etal~\cite{design_prism_sips_2012}, the design encodes the interval-ending position on the horizontal axis and the interval duration (window size) on the vertical axis.
The tip of the triangle (\subcref{fig:design}{A}) effectively summarizes the entire time series, with color encoding consistent with the correlation matrix.
The view provides a comprehensive overview of statistical aggregates across all possible subintervals at a glance.
Additionally, a line chart to the right complements the heatmap, displaying financial time series data with axes tailored to each stock's value range.

Contrary to using interval-starting positions~\cite{related_stock_pairwise_keim_2006} and midpoints~\cite{design_prism_sips_2012} on the horizontal axis, our interval-ending position layout is particularly suited to the finance domain for several reasons:
i) It matches the conventional practice of observing stocks at their closing prices, simplifying the learning process for users;
ii) It facilitates a natural progression of correlation tracking from the start date, clearly visualized along the triangle's slant and avoiding confusion from adjacent colors in a midpoint layout;
iii) It avoids the potential for misleading future predictions inherent in other layouts.
To optimize this layout, we use a 1D array with an indexing system $i \mapsto (x,y)$ to reduce rendering time and memory by half. Let $n$ denote the maximum window size, the mark's coordinates, having the origin placed at the top left corner and 0-indexed array, are $y = \lfloor(\sqrt{8*i+1}-1)/2\rfloor$ and $x = n - 1 - y + i - \lfloor(y*(y+1))/2\rfloor$. 

The view metaphorically resembles light's refraction through a prism to help users understand the patterns within time series data.
Users can interactively explore the data by brushing across the x-axis to zoom into different time frames, identifying two key visual patterns (\textbf{CV3}):
i) \textbf{vertical} lines (\subcref{fig:design}{B}) indicate significant daily impacts that disrupt and equalize previous trends, and 
ii) \textbf{slant} lines (\subcref{fig:design}{C}) mark the start of new long-term trends from the starting date.
For example, the prism (\cref{fig:design}) shows a noticeable trend beginning in April and abruptly impacted in mid-June.
Moreover, we provide several prisms for comparative analysis with the market indices (\subcref{fig:teaser}{D1}), the corresponding industry index (\subcref{fig:teaser}{D2}), and against another selected stock (\subcref{fig:teaser}{D3}).

%% file: 6-evaluation.tex
\section{Evaluation}

In this section, we outline the evaluation protocol to assess the effectiveness and usability of {\name} in identifying dynamic financial clusters.
We focus on customizing concepts based on the users' needs. 
We first introduce a usage scenario demonstrating how an ordinary investor can use {\name} to reveal valuable insights within the financial market swiftly. 
The first case is the material for the expert interviews to illustrate the functions of {\name}.
The second case is suggested by interviewee E1, who has discovered an intriguing result that one has not expected, providing a more comprehensive view of {\name}'s capabilities.

\textbf{Participants.} We interviewed three domain experts, E1-3, to evaluate {\name}'s usefulness and effectiveness. 
E1 is an experienced financial analyst who has worked in fundamental analysis and private equity sectors.
E1's working experience endorses his ability to give constructive advice toward interpreting fundamentals, which could be moderately translated to knowledge-driven exploration in {\name}.
E1 participated in the requirement formulation process and was invited to the expert interview to validate the requirements.
E2 is a certified actuarial analyst and a Chartered Financial Analyst (CFA) level III charter holder.
Specializing in quantitative methods in risk management, E2 is also well-equipped with financial analytical knowledge such as asset valuation and portfolio management. 
E3 is a senior investing consultant with over 20 years of experience in the banking industry and rich expert knowledge in the global market.
None of them are co-authors.

\textbf{Procedure.} 
The interview was conducted remotely. 
The {\name} system was deployed online for access. 
Initially, each expert was asked to fill out a consent form and a demographic questionnaire about their background. 
Then, we introduced our motivation, key concepts, framework, and visual designs of {\name} with an example-based introduction~\cite{yang2023examples}. 
To illustrate the system's functionality, we showcased the usage scenario in \cref{sec:usage_scenario}, lasting approximately 15 minutes. 
The experts then explored the system for about half an hour. 
They were encouraged to enter the stocks they were interested in and then analyze and generate a personalized concept. 
After that, we conducted a 20-minute semi-structured interview with experts, asking them to provide a detailed evaluation of our system, including its usability, effectiveness, and visual appeal. 
To conclude, we held a 10-minute open discussion, providing a forum for additional comments or suggestions.

\begin{figure*}[tb]
    \setlength{\belowcaptionskip}{-0.3cm}
	\centering
	\includegraphics[width=\linewidth]{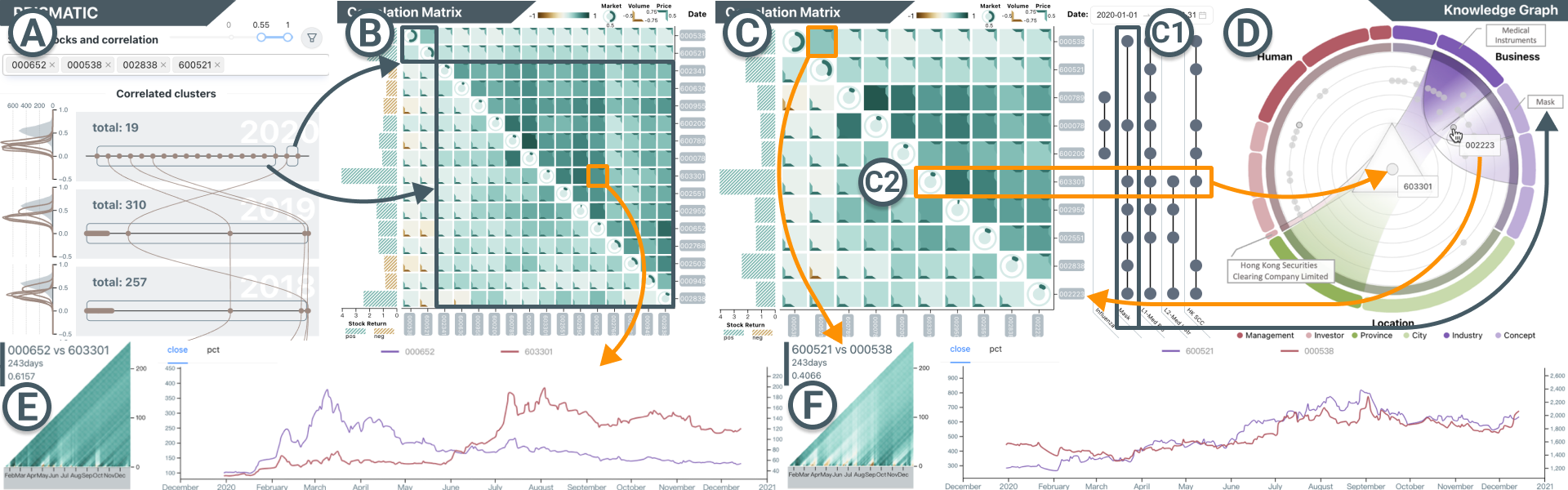}
        \caption{This usage scenario describes an analysis of four medical industry stocks during the 2020 COVID-19 outbreak. From a detailed analysis of the cluster related to `mask' and `influenza' concepts (B), the hidden insight behind correlations (diverging performance under high correlation (E) and converging performance under low correlation (F)) is exposed and guides the refinement the cluster. Besides, the knowledge graph view (D) facilitates the exploration of qualitative business proximity. The results are summarized (C) to validate the cluster based on historical performance and knowledge-driven insights.}
    \label{fig:case1}
\end{figure*}

\subsection{Usage scenario: medicine stocks during COVID-19}
\label{sec:usage_scenario}
During the COVID-19 outbreak, the global economy experienced turbulence, and the Chinese market saw a significant downturn after the Chinese New Year. 
However, due to its unique business nature, the medical industry has shown a divergent performance.

Richard, a newlywed investor aiming to learn from past events, selects four medical industry stocks from the internet to analyze their performance in 2020 (\subcref{fig:case1}{A}).
He wants to create his own concept stock for medicines to avoid taking too much risk for sudden events.
He focuses on two mask manufacturers (`000652,' `002838'), a renowned Chinese medicine company (`000538'), and a leading Western medicine company (`600521').
The distribution plots indicate very few companies maintained correlations over 0.5 with the selected stocks in 2020 (\textbf{CG1}).
Setting a 0.55 correlation threshold, Richard discovers only 19 stocks correlated with his selected four in 2020, but there were 310 and 257 in the previous years (\textbf{CG2}). 
He also notices that the two rectangles depicted two distinct components in 2020, suggesting that one of his selected stocks might have misaligned correlations.

Focusing on 2020, Richard observes a large and smaller cluster in the correlation matrix (\subcref{fig:case1}{B}, annotated with orange rectangular boxes).
He realizes that these two clusters correspond to the two components displayed in the previous view, prompting him to examine these clusters separately.
Starting with the larger cluster, he reviews stock performance in the first quarter 2020.
He notices that the matrix is less filled in the diagonal, meaning that the cluster's correlation with the volatile market index is close to zero.
He selects `000652,' a mask manufacturer,
He finds that the stock is associated with the `mask' and `influenza' concepts through the knowledge graph view (\textbf{CE1}, \subcref{fig:case1}{D}).
By incorporating these concepts into the UpSet plot, Richard finds most stocks in the cluster also linked to the `mask' concept (\textbf{CE2}, \subcref{fig:case1}{C1}).
Switching back to the 2020 overview, `603301' stands out for its exceptional performance in price return (\textbf{CV2}, \subcref{fig:case1}{C2}).

Interested in understanding the differences between `603301' and `000652,' Richard turns to the Prism time series view (\subcref{fig:case1}{E}). 
The correlation coefficient is 0.6157 for 243 trading days throughout the year.
The prism also shows uniform color without many deviations, indicating that the correlation is aligned across all sub-intervals.
Despite the initial appearance of a strong correlation, their performance patterns are quite different. 
The line chart reveals that `000652' has seen a spike in the first three months but declined after that.
On the contrary, `603301' has stroked a record high since June and reached a stagnation.
This observation cautions Richard about the speculative nature of concept stocks (\textbf{CV3}, \subcref{fig:case1}{E}), leading him to exclude `000652' from the concept.

Furthermore, Richard turns his attention to `603301,' exploring the knowledge graph view to discover potential correlated stocks (\textbf{CE1}, \subcref{fig:case1}{D}).
Within the multi-layer network, he navigates through three layers of \rw{business relational knowledge}. 
Upon selecting the `medical instruments' segment, which has the largest segment, stocks sharing this item appear as nodes. 
Since the proximity of a node to the center indicates a stronger relationship with `603301,' he clicks on the node `002223,' revealing its connections across other business, location, and human layers.
Five chords indicate a strong business proximity between the two selected stocks. 
Thus he double-clicks and adds the node to the cluster.

Richard examines the correlation matrix and time series, comparing the two stocks in detail. 
He also uses the UpSet plot and the bar chart for auxiliary information (\textbf{CV1}, \subcref{fig:case1}{C2}).
Through careful validation, he confirms that `603301' and `002223' demonstrate similar performance and are highly correlated. 
Consequently, it is clear that `002223' should remain within this cluster.
By continuing to explore and validate data-driven clusters with knowledge-driven similarity, he added and deleted some stocks to form the final concept stock cluster.
By continuing this iterative process of exploration and validation of data-driven clusters based on knowledge-driven similarity, he creates a bespoke concept stock cluster.

Additionally, Richard conducts a thorough review of the smaller cluster located in the top left corner of the matrix that includes a Chinese medicine company and a Western medicine firm.
Despite their modest correlation coefficient of 0.4, he notices a remarkable similarity in their stock trends (\subcref{fig:case1}{F}) for stock price and change in percentage.
This observation prompts Richard to keep these two medicine stocks within the cluster, attributing their retention to the evident strength of their comovement. 
After engaging in interactive clustering with {\name}, the unique set of concept stocks, specifically tailored to his requirements, is successfully formulated.

\subsection{Case study: media stocks}
\label{sec:case_study}
\begin{figure}[tb]
    \setlength{\belowcaptionskip}{-0.3cm}
	\centering
	\includegraphics[width=\linewidth]{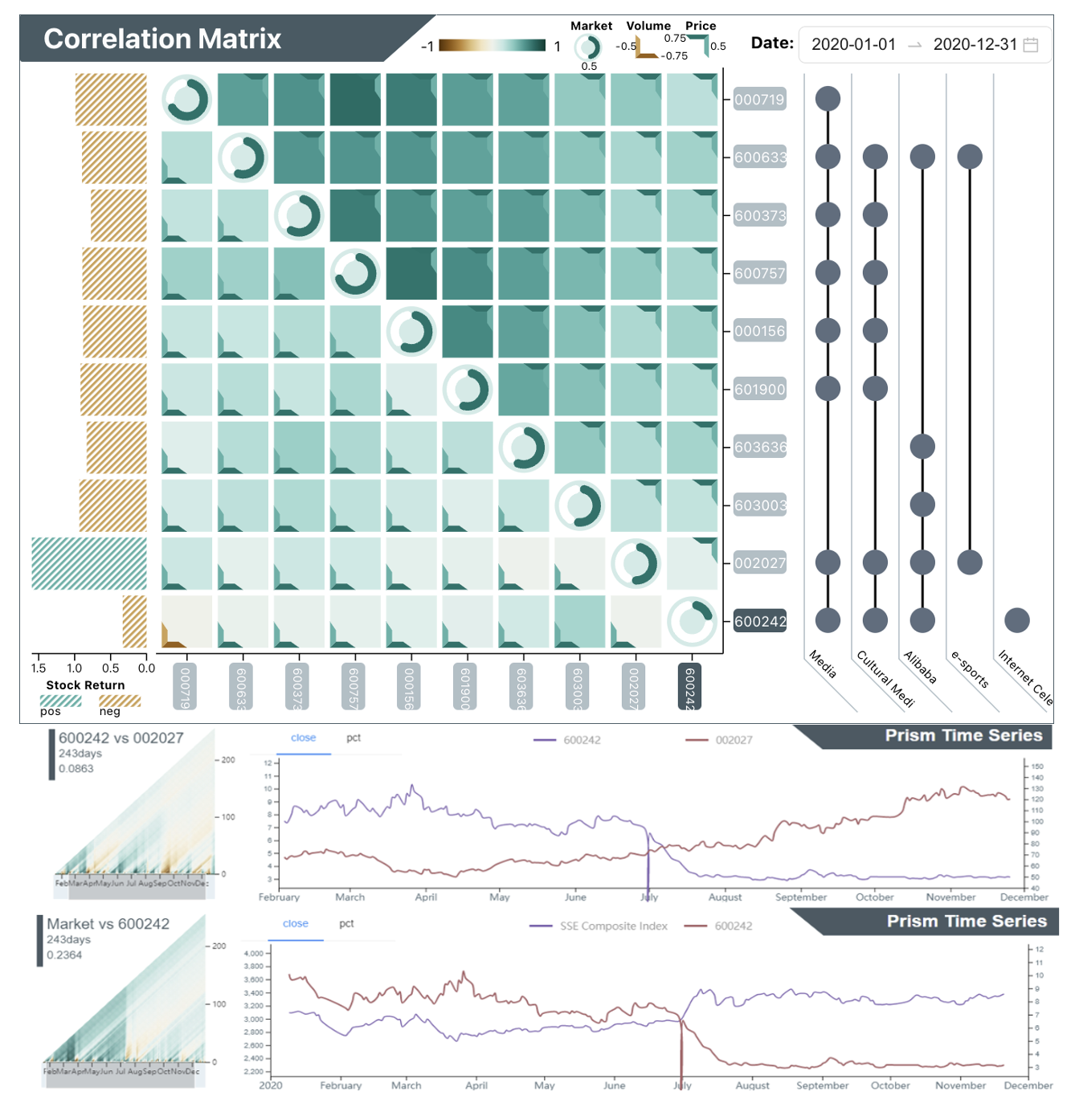}
    \caption{{\name} provides insights behind the correlations and trends of media stocks `600373' and `002027,' depicted here, leading to informed stock selections and portfolio optimizations based on in-depth data and \rw{business relational knowledge} exploration.}
    \label{fig:case2}
\end{figure}

This case (\cref{fig:case2}) shows that {\name} is an invaluable tool for E1, allowing for in-depth data exploration and discovering unique insights that might otherwise remain hidden.
As an experienced practitioner, E1 interacts with {\name} in a highly targeted manner and conducts a detailed analysis of two media industry stocks: `600373' and `002027.'
She sets a correlation threshold of 0.63 and discovers that these stocks have few correlated stocks between 2017 and 2019 (\textbf{CG1}, \textbf{CG2}). 
She identifies a highly correlated cluster in 2020 with a high correlation with the market index, represented by a green diagonal.
Except for `002027,' all stocks have significant losses in their stock returns (\textbf{CV2}).
This outcome doesn't surprise E1, given her familiarity with the media sector's dynamics and acknowledgment of the trend.

E1's exploration continues with the stock `600373,' leveraging the multi-layer network to obtain related \rw{business relational knowledge}.
By clicking on the outer segment of the business layer, the `media' industry and `cultural media' concept are added to the UpSet plot.
Similarly, E1 explores `002027' and filters for the `Alibaba' concept. 
Four stocks with shared relationships are identified and added to the cluster (\textbf{CE1}).

E1's attention is drawn to `600242,' distinguished by its unique returns despite sharing knowledge items with `002027,' such as involvement in `media,' `cultural media,' and `Alibaba' concepts (\textbf{CV1}).
Initial observations in the correlation matrix report that the colors on the intersection between `600242' and `002027' are very light and close to white, suggesting their negligible correlations in both price and volume.
For a more detailed quantitative analysis, E1 examines the prism time series view. 
The pairwise prism confirms the weak correlation of merely 0.0863, revealing a stark contrast in their trends, with `002027' ascending and `600242' descending all the time.
The prism against the market shows that the company performs in line with the market index until a pivotal change in July (\textbf{CV3}).
Drilling down to the phenomenon behind the observed vertical line, E1 identifies a critical moment when `600242' experienced a suspension and subsequent significant decline.

Seeking to understand the qualitative factors behind these trends, E1 switches to the knowledge graph view to get a complementary analysis. 
After examining the business layer, E1 discovers the concepts of `e-sports' and `Internet celebrity economy' and believes they are the reasons behind the diverging performance.
Adding to the UpSet plot, `002027' is linked to `e-sports,' while `600242' is associated with the `Internet celebrity economy.'
Equipped with comprehensive correlation and \rw{business relational knowledge} analysis, E1 retains the profitable `002027' and excludes `600242' from the portfolio.
The analysis helps E1 develop an investment preference in e-sports over Internet celebrities in 2020. 
After the interview, this decision is justified as E1 finds that `600242' faced a debt crisis, highlighting the value of {\name} in aiding professionals with stock selection.

\subsection{Expert interview}
We present a summary of semi-structured interviews and open discussions with three financial experts, focusing on their feedback regarding {\name}'s usability, its comparison with other systems, the transferability of its knowledge-driven approach, the learning curve for visualizations, the design aspects, and their suggestions for enhancements. 
Their collective feedback highlights the tool's applicability and effectiveness in financial markets analysis.

\textbf{Usability.} 
All three experts affirmed {\name} as a valuable tool for a diverse user base, including professional fund managers, buy-side financial analysts, researchers, and individual investors. 
They recognized its capacity to facilitate a deeper understanding of the market and specific securities.
After introducing visualizations and case studies, they understood each view's workflow and corresponding functions clearly, describing the approach as ``clear and coherent'' and making sense.
E1 appreciated the tool's capability to demystify the nuanced concept of `concept stocks,' ``\textit{It varies significantly across brokers and institutions, but {\name} is rooted in data.}''
This data-driven approach achieves a higher level of precision and personalization rarely seen in traditional portfolio construction practices, which typically depend on observing different financial indicators before cross-referencing with business characteristics in commercial databases.
E2 also remarked that {\name}'s approach to integrating unstructured data into a structured analysis framework is considerably improved over traditional methods, as `knowledge' is often unstructured in the financial context.

\textbf{Comparison with other systems.}
The experts appreciated that {\name} offers standout features, particularly its advanced visualization capabilities, which distinguish it from basic tools like Excel, which are still prevalent in many institutions.
The system's variety of chart types supports an in-depth visual analysis to derive insights from multiple data perspectives, unlike popular platforms such as \textit{iFinD} and \textit{Great Wisdom}, which primarily focus on mere data presentation without substantial analytical depth. 
These platforms provide information about concept stocks but fail to offer insights into the logic behind their classifications. 
In contrast, {\name} provides an analytical framework to craft users' own concept stocks, including cluster generation, exploration, and validation.
E1 commented, ``\textit{The tools we use today are still in the 1.0 Stone Age, but {\name} has reached the 3.0 era.}''

Additionally, E3 highlighted the potential of {\name} to refine and enhance financial clusters from existing platforms, tailoring them more closely to specific user requirements. 
The conventional process is often updated slowly due to the need for validation and concerns over corporate image. 
The rigid clusters in existing systems rely heavily on the subjective judgment of the financial research firm's senior managers.
{\name}'s capability to quickly adjust portfolios based on real-time data analysis was recognized as a crucial advantage that sets it distinctly ahead of its counterparts.

\textbf{Transferability.}
The relevance and flexibility of {\name}'s knowledge layers were highlighted as key strengths, allowing for their application across various scenarios beyond traditional financial market analysis.
E1 praised the system for its ``one-stop clustering approach,'' which streamlines financial data analysis by integrating disparate data sources into a cohesive analysis framework.
E3 expressed interest in using this workflow to detect anomalies, such as the unusual divergence in correlation among stocks within the same sector, which often signals potential for excess returns.
``The combination of knowledge graph and prism time series view helps reveal such disparities,'' E3 noted.

From a quantitative analyst's perspective, E2 recommended incorporating customizable metrics to suit different analytical objectives.
For example, analysts focus on portfolio's alpha for overall returns and conditional value-at-risk for tail risks.
Providing these metrics other than correlation supports a more comprehensive quantitative analysis.
From an actuarial perspective, E2 expressed interest in applying a similar interactive clustering workflow to insurance policy and contract analysis, where identifying clusters of policyholders based on shared characteristics is crucial. 
For example, the dimensions in the current multi-layer network can be replaced by ``health history, financial condition, and life status,'' giving high praise to the transferability of the approach.

\textbf{Learning curve.} 
Feedback from experts on the learning curve associated with {\name} varied, highlighting the system's overall user-friendly design despite some initial challenges with specific visualizations, such as the correlation matrix.
E3 mentioned that the correlation matrix requires a shift from traditional number-reading practices, although the heatmap design compensates by enhancing cluster identification.
E1 acknowledged the complexity of the Prism view but noted that the prism metaphor significantly aids in understanding the concerned visual patterns.
E2 pointed out the complexity of navigating the knowledge graph view due to its dense visual information but appreciated the interactive nature to ease the learning process.
While it is challenging to pinpoint which visualization has the steepest learning curve, an individual's professional background and working habits influence their ease of adapting to {\name}'s visualizations.

\textbf{Visualization designs.} 
The aesthetic and functional aspects of {\name}'s design received high praise for enhanced user experience, with experts appreciating the thoughtful use of color and hierarchical clustering to guide analytical focus.
E2 found the matrix design intuitive and remarkably engaging: ``It is quite refreshing and pleasing to read.''
This preference underscores the balance between aesthetics and functionality, with the legend-aside design optimizing space for additional data insights. 
E1 remarked on the Prism view's superior detail, stating that reading statistics is ``nowhere near'' as informative.
All experts especially praised the prism view's innovative summarization power, enabling efficient comparison of financial impacts across different timescales.

\textbf{Suggestions.}
Experts expressed interest in seeing {\name} expand its capabilities to include inter-cluster comparisons among user-defined concepts. 
This could further enrich financial analysis by comparing broader ranges of asset sub-classes and putting analysis results into practice.
E3 suggested a shift from merely filtering to actively adding business tags in the cluster generation, streamlining the direct observation of correlations. 
Visualizing the process behind multi-view clustering was also requested to offer users insights into the cluster generation and exploration mechanism. 
This indicates a desire for even more granular control and understanding of the analytical process.

%% file: 7-discussion.tex
\section{Discussion}

\textbf{Complementing quantitative and qualitative analysis.}
Quantitative analysis relies on numerical data and time series for investment decisions and often utilizes correlation for simplicity. 
However, this approach may oversimplify the underlying complex relationships, leading to incorrect causation assumptions.
This is particularly true for highly correlated stocks or with smaller market values, as they often follow the grand market fluctuations.
Our approach advocates a balanced integration of quantitative and qualitative analysis to craft bespoke financial clusters, introducing a nuanced perspective that considers broader market influences and individual stock behaviors. 
By leveraging \rw{business relational knowledge} alongside quantitative data, {\name} provides a more reasonable interpretation of correlations, balancing subjective insights with empirical evidence.

\textbf{Deep reasoning with concept stocks.}
Current financial analysis tools and platforms often present concept stocks ``as-is'' without explaining the underlying logic or clustering criteria, leaving users to navigate these concepts without guidance.
{\name} empowers users to define their personalized concept stocks, anchored in data- and knowledge-driven decision factors.
This adaptability becomes particularly valuable in a volatile market, enabling users to adjust their strategies based on the latest information. 
{\name} is designed with the understanding that investment strategies are highly personal based on the individual's knowledge base, risk preference, expertise, and financial goals. 
Its interactive clustering process not only provides a deeper understanding of why certain stocks belong together but also employs a bottom-up approach based on outlier detection, enabling users to refine their clusters methodically. 
Making the logic behind multi-view clustering transparent aligns with a human-in-the-loop workflow, enhancing the rationale of concept stocks and uncovering underlying associative factors that drive stock correlations.

\textbf{Towards responsible interpretation.} 
While {\name} utilizes keywords to represent business relational knowledge, there is potential for adopting algorithms that provide explanations in more intuitive, natural language.
The advent of Large Language Models (LLMs) has offered such opportunities for extracting insights from unstructured information.
When harnessed correctly, these LLMs enhance decision-making with their ability to parse and rationalize financial narratives.
However, the reliability of LLM-generated content necessitates rigorous validation for high-stakes scenarios.
Refined prompts can improve output accuracy~\cite{feng2024promptmagician}, but incorporating quantitative analyses is crucial for mitigating potential hallucinations.
Given the black-box nature of LLMs and the latency in their knowledge update cycles, our integrated approach, which validates qualitative knowledge with quantitative data, offers a potential safeguard against these pitfalls.
By upholding this balance, we not only maintain the integrity of the insights derived from LLMs but also establish a level of transparency and control critical for retaining confidence in the models' recommendations~\cite{feng2023xnli}, especially in the rapidly evolving landscape of financial markets.

\subsection{Limitations}
Our experiment indicates a noticeable performance drop in rendering the prism view and correlation matrix for clusters with over 40 elements due to the limitations of our D3.js~\cite{d3} implementation. 
Switching to alternative rendering techniques like Canvas could improve performance significantly, especially since the design does not require dynamic interactions.
\rc{Additionally, we acknowledge that the visual interface could be further enhanced by incorporating more comprehensive guidance to bridge the knowledge gap between system goals and analytical insights~\cite{discussion_guidance_2020}.}

The analysis of time-sensitive market dynamics relies primarily on quantitative analysis, \ie stock prices and trading volumes, as the relational knowledge in {\name} is static.
However, integrating dynamic knowledge that tracks significant events' timing and sequential patterns~\cite{he2024videopro} could offer more precise real-time market analyses. 
This would enable users to better understand and react to market volatility.
Designing visualization techniques to represent dynamic knowledge remains an open challenge.

{\name}'s effectiveness has been evaluated primarily through expert interviews, reflecting a credibility gap.
However, tools that integrate qualitative and quantitative analyses are scarce, leading to a lack of baseline systems.
While many platforms offer up-to-the-minute data, they fall short of facilitating in-depth analytical reasoning. 
Future research could mitigate this gap by designing task-based evaluations and developing ablated systems to evaluate the impact of combining qualitative and quantitative data on analysis outcomes.

%% file: 8-conclusion.tex
\section{Conclusion}

We propose {\name} to integrate quantitative and qualitative analysis for identifying concept stocks.
We formulate a set of design requirements for the financial context and, accordingly, design a visual interface to let users enrich the data-driven clustering result with knowledge-driven clustering exploration.
Three levels of analysis tasks were generalized to better understand the financial market clusters. 
Our system was \rw{evaluated qualitatively by two representative case studies and an expert interview afterward.} 
The result indicates that {\name} performs efficiently in financial correlation cluster exploration and concept stock analysis.


%% file: main_tvcg.bbl
\begin{thebibliography}{10}

\bibitem{related_stock_pairwise_alsakran_2010}
J.~Alsakran, Y.~Zhao, and X.~Zhao.
\newblock {Tile-based Parallel Coordinates and its Application in Financial Visualization}.
\newblock In {\em Visualization and Data Analysis 2010}, vol. 7530, pp. 21--32. SPIE, 2010. \href{https://doi.org/10.1117/12.838819}
{doi: {{%
10\hspace{.1pt}\discretionary{.}{%
}{.}\hspace{.4pt}1117\discretionary{/}{%
}{/}12\hspace{.1pt}\discretionary{.}{%
}{.}\hspace{.4pt}838819}}}


\bibitem{motivation_arleo_2023}
A.~Arleo, C.~Tsigkanos, R.~A. Leite, S.~Dustdar, S.~Miksch, and J.~Sorger.
\newblock {Visual Exploration of Financial Data with Incremental Domain Knowledge}.
\newblock {\em Computer Graphics Forum}, 42(1):101--116, 2023. \href{https://doi.org/10.1111/cgf.14723}
{doi: {{%
10\hspace{.1pt}\discretionary{.}{%
}{.}\hspace{.4pt}1111\discretionary{/}{%
}{/}cgf\hspace{.1pt}\discretionary{.}{%
}{.}\hspace{.4pt}14723}}}


\bibitem{system_correlation_aste_2010}
T.~Aste, W.~Shaw, and T.~D. Matteo.
\newblock {Correlation Structure and Dynamics in Volatile Markets}.
\newblock {\em New Journal of Physics}, 12(8):085009, Aug. 2010. \href{https://doi.org/10.1088/1367-2630/12/8/085009}
{doi: {{%
10\hspace{.1pt}\discretionary{.}{%
}{.}\hspace{.4pt}1088\discretionary{/}{%
}{/}1367\discretionary{%
}{-}{-}2630\discretionary{/}{%
}{/}12\discretionary{/}{%
}{/}8\discretionary{/}{%
}{/}085009}}}


\bibitem{related_clustering_bae_2020}
J.~Bae, T.~Helldin, M.~Riveiro, S.~Nowaczyk, M.-R. Bouguelia, and G.~Falkman.
\newblock {Interactive Clustering: a Comprehensive Review}.
\newblock {\em ACM Computing Surveys (CSUR)}, 53(1):1--39, 2020. \href{https://doi.org/10.1145/3340960}
{doi: {{%
10\hspace{.1pt}\discretionary{.}{%
}{.}\hspace{.4pt}1145\discretionary{/}{%
}{/}3340960}}}


\bibitem{design_matrix_ball_1969}
R.~Ball and P.~Brown.
\newblock {Portfolio Theory and Accounting}.
\newblock {\em Journal of Accounting Research}, 7(2):300--323, 1969. \href{https://doi.org/10.2307/2489972}
{doi: {{%
10\hspace{.1pt}\discretionary{.}{%
}{.}\hspace{.4pt}2307\discretionary{/}{%
}{/}2489972}}}


\bibitem{related_network_basole_2013}
R.~C. {Basole}, T.~{Clear}, M.~{Hu}, H.~{Mehrotra}, and J.~{Stasko}.
\newblock {Understanding Interfirm Relationships in Business Ecosystems with Interactive Visualization}.
\newblock {\em IEEE Transactions on Visualization and Computer Graphics}, 19(12):2526--2535, 2013. \href{https://doi.org/10.1145/3185047}
{doi: {{%
10\hspace{.1pt}\discretionary{.}{%
}{.}\hspace{.4pt}1145\discretionary{/}{%
}{/}3185047}}}


\bibitem{related_network_basole_2017}
R.~C. Basole, T.~Major, and A.~Srinivasan.
\newblock {Understanding Alliance Portfolios Using Visual Analytics}.
\newblock {\em ACM Trans. Manage. Inf. Syst.}, 8(4), Aug. 2017. \href{https://doi.org/10.1145/3086308}
{doi: {{%
10\hspace{.1pt}\discretionary{.}{%
}{.}\hspace{.4pt}1145\discretionary{/}{%
}{/}3086308}}}


\bibitem{related_network_basole_2018}
R.~C. Basole, A.~Srinivasan, H.~Park, and S.~Patel.
\newblock {Ecoxight: Discovery, Exploration, and Analysis of Business Ecosystems Using Interactive Visualization}.
\newblock {\em ACM Trans. Manage. Inf. Syst.}, 9(2), Apr. 2018. \href{https://doi.org/10.1145/3185047}
{doi: {{%
10\hspace{.1pt}\discretionary{.}{%
}{.}\hspace{.4pt}1145\discretionary{/}{%
}{/}3185047}}}


\bibitem{design_network_beck_2017}
F.~Beck, M.~Burch, S.~Diehl, and D.~Weiskopf.
\newblock {A Taxonomy and Survey of Dynamic Graph Visualization}.
\newblock {\em Computer Graphics Forum}, 36(1):133--159, 2017. \href{https://doi.org/10.1111/cgf.12791}
{doi: {{%
10\hspace{.1pt}\discretionary{.}{%
}{.}\hspace{.4pt}1111\discretionary{/}{%
}{/}cgf\hspace{.1pt}\discretionary{.}{%
}{.}\hspace{.4pt}12791}}}


\bibitem{d3}
M.~Bostock, V.~Ogievetsky, and J.~Heer.
\newblock {D³ Data-Driven Documents}.
\newblock {\em IEEE Transactions on Visualization and Computer Graphics}, 17(12):2301--2309, 2011. \href{https://doi.org/10.1109/TVCG.2011.185}
{doi: {{%
10\hspace{.1pt}\discretionary{.}{%
}{.}\hspace{.4pt}1109\discretionary{/}{%
}{/}TVCG\hspace{.1pt}\discretionary{.}{%
}{.}\hspace{.4pt}2011\hspace{.1pt}\discretionary{.}{%
}{.}\hspace{.4pt}185}}}


\bibitem{related_clustering_cao_2011}
N.~Cao, D.~Gotz, J.~Sun, and H.~Qu.
\newblock {DICON: Interactive Visual Analysis of Multidimensional Clusters}.
\newblock {\em IEEE Transactions on Visualization and Computer Graphics}, 17(12):2581--2590, 2011. \href{https://doi.org/10.1109/TVCG.2011.188}
{doi: {{%
10\hspace{.1pt}\discretionary{.}{%
}{.}\hspace{.4pt}1109\discretionary{/}{%
}{/}TVCG\hspace{.1pt}\discretionary{.}{%
}{.}\hspace{.4pt}2011\hspace{.1pt}\discretionary{.}{%
}{.}\hspace{.4pt}188}}}


\bibitem{related_clustering_cao_2010}
N.~Cao, J.~Sun, Y.-R. Lin, D.~Gotz, S.~Liu, and H.~Qu.
\newblock {FacetAtlas: Multifaceted Visualization for Rich Text Corpora}.
\newblock {\em IEEE Transactions on Visualization and Computer Graphics}, 16(6):1172--1181, 2010. \href{https://doi.org/10.1109/TVCG.2010.154}
{doi: {{%
10\hspace{.1pt}\discretionary{.}{%
}{.}\hspace{.4pt}1109\discretionary{/}{%
}{/}TVCG\hspace{.1pt}\discretionary{.}{%
}{.}\hspace{.4pt}2010\hspace{.1pt}\discretionary{.}{%
}{.}\hspace{.4pt}154}}}


\bibitem{related_clustering_cavallo_2019}
M.~{Cavallo} and \c{c}. {Demiralp}.
\newblock {Clustrophile 2: Guided Visual Clustering Analysis}.
\newblock {\em IEEE Transactions on Visualization and Computer Graphics}, 25(1):267--276, 2019. \href{https://doi.org/10.1109/TVCG.2018.2864477}
{doi: {{%
10\hspace{.1pt}\discretionary{.}{%
}{.}\hspace{.4pt}1109\discretionary{/}{%
}{/}TVCG\hspace{.1pt}\discretionary{.}{%
}{.}\hspace{.4pt}2018\hspace{.1pt}\discretionary{.}{%
}{.}\hspace{.4pt}2864477}}}


\bibitem{discussion_guidance_2020}
D.~Ceneda, N.~Andrienko, G.~Andrienko, T.~Gschwandtner, S.~Miksch, N.~Piccolotto, T.~Schreck, M.~Streit, J.~Suschnigg, and C.~Tominski.
\newblock {Guide Me in Analysis: A Framework for Guidance Designers}.
\newblock {\em Computer Graphics Forum}, 39(6):269--288, 2020. \href{https://doi.org/10.1111/cgf.14017}
{doi: {{%
10\hspace{.1pt}\discretionary{.}{%
}{.}\hspace{.4pt}1111\discretionary{/}{%
}{/}cgf\hspace{.1pt}\discretionary{.}{%
}{.}\hspace{.4pt}14017}}}


\bibitem{chen2024fmlens}
L.~Chen, C.~Cheng, H.~Wang, X.~Wang, Y.~Tian, X.~Yue, W.~Kam-Kwai, H.~Zhang, S.~Hong, and Q.~Li.
\newblock {FMLens: Towards Better Scaffolding the Process of Fund Manager Selection in Fund Investments}.
\newblock {\em IEEE Transactions on Visualization and Computer Graphics}, pp. 1--17, 2024. \href{https://doi.org/10.1109/TVCG.2024.3394745}
{doi: {{%
10\hspace{.1pt}\discretionary{.}{%
}{.}\hspace{.4pt}1109\discretionary{/}{%
}{/}TVCG\hspace{.1pt}\discretionary{.}{%
}{.}\hspace{.4pt}2024\hspace{.1pt}\discretionary{.}{%
}{.}\hspace{.4pt}3394745}}}


\bibitem{system_cluster_chen_2022}
M.-S. Chen, J.-Q. Lin, X.-L. Li, B.-Y. Liu, C.-D. Wang, D.~Huang, and J.-H. Lai.
\newblock {Representation Learning in Multi-view Clustering: A Literature Review}.
\newblock {\em Data Science and Engineering}, 7(3):225--241, 2022. \href{https://doi.org/10.1007/s41019-022-00190-8}
{doi: {{%
10\hspace{.1pt}\discretionary{.}{%
}{.}\hspace{.4pt}1007\discretionary{/}{%
}{/}s41019\discretionary{%
}{-}{-}022\discretionary{%
}{-}{-}00190\discretionary{%
}{-}{-}8}}}


\bibitem{chen2024ethereum}
X.~Chen, X.~Zhang, Z.~Wang, K.~Yu, W.~Kam-Kwai, H.~Guo, and S.~Chen.
\newblock {Visual Analytics for Security Threats Detection in Ethereum Consensus Layer}.
\newblock {\em Journal of Visualization}, 27:469--483, Mar. 2024. \href{https://doi.org/10.1007/s12650-024-00969-z}
{doi: {{%
10\hspace{.1pt}\discretionary{.}{%
}{.}\hspace{.4pt}1007\discretionary{/}{%
}{/}s12650\discretionary{%
}{-}{-}024\discretionary{%
}{-}{-}00969\discretionary{%
}{-}{-}z}}}


\bibitem{motivation_chen_2018}
Y.~Chen, Z.~Wei, and X.~Huang.
\newblock {Incorporating Corporation Relationship via Graph Convolutional Neural Networks for Stock Price Prediction}.
\newblock In {\em Proc. CIKM}, pp. 1655--1658. ACM, New York, 2018. \href{https://doi.org/10.1145/3269206.3269269}
{doi: {{%
10\hspace{.1pt}\discretionary{.}{%
}{.}\hspace{.4pt}1145\discretionary{/}{%
}{/}3269206\hspace{.1pt}\discretionary{.}{%
}{.}\hspace{.4pt}3269269}}}


\bibitem{related_clustering_das_2021}
S.~Das, B.~Saket, B.~C. Kwon, and A.~Endert.
\newblock {Geono-Cluster: Interactive Visual Cluster Analysis for Biologists}.
\newblock {\em IEEE Transactions on Visualization and Computer Graphics}, 27(12):4401--4412, 2021. \href{https://doi.org/10.1109/TVCG.2020.3002166}
{doi: {{%
10\hspace{.1pt}\discretionary{.}{%
}{.}\hspace{.4pt}1109\discretionary{/}{%
}{/}TVCG\hspace{.1pt}\discretionary{.}{%
}{.}\hspace{.4pt}2020\hspace{.1pt}\discretionary{.}{%
}{.}\hspace{.4pt}3002166}}}


\bibitem{motivation_deng_2019}
S.~Deng, N.~Zhang, W.~Zhang, J.~Chen, J.~Z. Pan, and H.~Chen.
\newblock {Knowledge-Driven Stock Trend Prediction and Explanation via Temporal Convolutional Network}.
\newblock In {\em Proc. WWW}. ACM, New York, 2019. \href{https://doi.org/10.1145/3308560.3317701}
{doi: {{%
10\hspace{.1pt}\discretionary{.}{%
}{.}\hspace{.4pt}1145\discretionary{/}{%
}{/}3308560\hspace{.1pt}\discretionary{.}{%
}{.}\hspace{.4pt}3317701}}}


\bibitem{related_knowledge_eirich_2022}
J.~Eirich, J.~Bonart, D.~Jäckle, M.~Sedlmair, U.~Schmid, K.~Fischbach, T.~Schreck, and J.~Bernard.
\newblock {IRVINE: A Design Study on Analyzing Correlation Patterns of Electrical Engines}.
\newblock {\em IEEE Transactions on Visualization and Computer Graphics}, 28(1):11--21, 2022. \href{https://doi.org/10.1109/TVCG.2021.3114797}
{doi: {{%
10\hspace{.1pt}\discretionary{.}{%
}{.}\hspace{.4pt}1109\discretionary{/}{%
}{/}TVCG\hspace{.1pt}\discretionary{.}{%
}{.}\hspace{.4pt}2021\hspace{.1pt}\discretionary{.}{%
}{.}\hspace{.4pt}3114797}}}


\bibitem{system_correlation_eisinga_2013}
R.~Eisinga, M.~t. Grotenhuis, and B.~Pelzer.
\newblock {The Reliability of a Two-Item Scale: Pearson, Cronbach, or Spearman-Brown?}
\newblock {\em International Journal of Public Health}, 58(4):637--642, Aug. 2013. \href{https://doi.org/10.1007/s00038-012-0416-3}
{doi: {{%
10\hspace{.1pt}\discretionary{.}{%
}{.}\hspace{.4pt}1007\discretionary{/}{%
}{/}s00038\discretionary{%
}{-}{-}012\discretionary{%
}{-}{-}0416\discretionary{%
}{-}{-}3}}}


\bibitem{related_knowledge_federico_2017}
P.~Federico, M.~Wagner, A.~Rind, A.~Amor-Amorós, S.~Miksch, and W.~Aigner.
\newblock {The Role of Explicit Knowledge: A Conceptual Model of Knowledge-Assisted Visual Analytics}.
\newblock In {\em Proc. VAST}, pp. 92--103. IEEE, 2017. \href{https://doi.org/10.1109/VAST.2017.8585498}
{doi: {{%
10\hspace{.1pt}\discretionary{.}{%
}{.}\hspace{.4pt}1109\discretionary{/}{%
}{/}VAST\hspace{.1pt}\discretionary{.}{%
}{.}\hspace{.4pt}2017\hspace{.1pt}\discretionary{.}{%
}{.}\hspace{.4pt}8585498}}}


\bibitem{motivation_feng_2019}
F.~Feng, X.~He, X.~Wang, C.~Luo, Y.~Liu, and T.-S. Chua.
\newblock {Temporal Relational Ranking for Stock Prediction}.
\newblock {\em ACM Trans. Inf. Syst.}, 37(2):1--30, Mar. 2019. \href{https://doi.org/10.1145/3309547}
{doi: {{%
10\hspace{.1pt}\discretionary{.}{%
}{.}\hspace{.4pt}1145\discretionary{/}{%
}{/}3309547}}}


\bibitem{feng2024promptmagician}
Y.~Feng, X.~Wang, W.~Kam-Kwai, S.~Wang, Y.~Lu, M.~Zhu, B.~Wang, and W.~Chen.
\newblock {PromptMagician: Interactive Prompt Engineering for Text-to-Image Creation}.
\newblock {\em IEEE Transactions on Visualization and Computer Graphics}, 30(1):295--305, 2024. \href{https://doi.org/10.1109/TVCG.2023.3327168}
{doi: {{%
10\hspace{.1pt}\discretionary{.}{%
}{.}\hspace{.4pt}1109\discretionary{/}{%
}{/}TVCG\hspace{.1pt}\discretionary{.}{%
}{.}\hspace{.4pt}2023\hspace{.1pt}\discretionary{.}{%
}{.}\hspace{.4pt}3327168}}}


\bibitem{feng2023xnli}
Y.~Feng, X.~Wang, B.~Pan, W.~Kam-Kwai, Y.~Ren, S.~Liu, Z.~Yan, Y.~Ma, H.~Qu, and W.~Chen.
\newblock {XNLI: Explaining and Diagnosing NLI-based Visual Data Analysis}.
\newblock {\em IEEE Transactions on Visualization and Computer Graphics}, pp. 1--14, 2023. \href{https://doi.org/10.1109/TVCG.2023.3240003}
{doi: {{%
10\hspace{.1pt}\discretionary{.}{%
}{.}\hspace{.4pt}1109\discretionary{/}{%
}{/}TVCG\hspace{.1pt}\discretionary{.}{%
}{.}\hspace{.4pt}2023\hspace{.1pt}\discretionary{.}{%
}{.}\hspace{.4pt}3240003}}}


\bibitem{introduction_fenn_2011}
D.~J. Fenn, M.~A. Porter, S.~Williams, M.~McDonald, N.~F. Johnson, and N.~S. Jones.
\newblock {Temporal Evolution of Financial-Market Correlations}.
\newblock {\em Phys. Rev. E}, 84:026109, Aug. 2011. \href{https://doi.org/10.1103/PhysRevE.84.026109}
{doi: {{%
10\hspace{.1pt}\discretionary{.}{%
}{.}\hspace{.4pt}1103\discretionary{/}{%
}{/}PhysRevE\hspace{.1pt}\discretionary{.}{%
}{.}\hspace{.4pt}84\hspace{.1pt}\discretionary{.}{%
}{.}\hspace{.4pt}026109}}}


\bibitem{related_stock_survey_flood_2016}
M.~D. Flood, V.~L. Lemieux, M.~Varga, and B.~{William Wong}.
\newblock {The Application of Visual Analytics to Financial Stability Monitoring}.
\newblock {\em Journal of Financial Stability}, 27:180--197, 2016. \href{https://doi.org/10.1016/j.jfs.2016.01.006}
{doi: {{%
10\hspace{.1pt}\discretionary{.}{%
}{.}\hspace{.4pt}1016\discretionary{/}{%
}{/}j\hspace{.1pt}\discretionary{.}{%
}{.}\hspace{.4pt}jfs\hspace{.1pt}\discretionary{.}{%
}{.}\hspace{.4pt}2016\hspace{.1pt}\discretionary{.}{%
}{.}\hspace{.4pt}01\hspace{.1pt}\discretionary{.}{%
}{.}\hspace{.4pt}006}}}


\bibitem{system_betweenness_freeman_1977}
L.~C. Freeman.
\newblock {A Set of Measures of Centrality Based on Betweenness}.
\newblock {\em Sociometry}, 40(1):35--41, 1977. \href{https://doi.org/10.2307/3033543}
{doi: {{%
10\hspace{.1pt}\discretionary{.}{%
}{.}\hspace{.4pt}2307\discretionary{/}{%
}{/}3033543}}}


\bibitem{task_gleicher_2018}
M.~{Gleicher}.
\newblock {Considerations for Visualizing Comparison}.
\newblock {\em IEEE Transactions on Visualization and Computer Graphics}, 24(1):413--423, 2018. \href{https://doi.org/10.1109/TVCG.2017.2744199}
{doi: {{%
10\hspace{.1pt}\discretionary{.}{%
}{.}\hspace{.4pt}1109\discretionary{/}{%
}{/}TVCG\hspace{.1pt}\discretionary{.}{%
}{.}\hspace{.4pt}2017\hspace{.1pt}\discretionary{.}{%
}{.}\hspace{.4pt}2744199}}}


\bibitem{system_data_guo_2018}
X.~Guo, H.~Zhang, and T.~Tian.
\newblock {Development of Stock Correlation Networks using Mutual Information and Financial Big Data}.
\newblock {\em PLOS ONE}, 13(4):1--16, Apr. 2018. \href{https://doi.org/10.1371/journal.pone.0195941}
{doi: {{%
10\hspace{.1pt}\discretionary{.}{%
}{.}\hspace{.4pt}1371\discretionary{/}{%
}{/}journal\hspace{.1pt}\discretionary{.}{%
}{.}\hspace{.4pt}pone\hspace{.1pt}\discretionary{.}{%
}{.}\hspace{.4pt}0195941}}}


\bibitem{introduction_gupta_1972}
M.~C. Gupta and R.~J. Huefner.
\newblock {A Cluster Analysis Study of Financial Ratios and Industry Characteristics}.
\newblock {\em Journal of Accounting Research}, pp. 77--95, 1972. \href{https://doi.org/10.1109/10.2307/2490219}
{doi: {{%
10\hspace{.1pt}\discretionary{.}{%
}{.}\hspace{.4pt}1109\discretionary{/}{%
}{/}10\hspace{.1pt}\discretionary{.}{%
}{.}\hspace{.4pt}2307\discretionary{/}{%
}{/}2490219}}}


\bibitem{background_cluster_gwilym_2016}
O.~A. Gwilym, I.~Hasan, Q.~Wang, and R.~Xie.
\newblock {In Search of Concepts: The Eﬀects of Speculative Demand on Stock Returns}.
\newblock {\em European Financial Management}, 22(3):427--449, 2016. \href{https://doi.org/10.1111/eufm.12067}
{doi: {{%
10\hspace{.1pt}\discretionary{.}{%
}{.}\hspace{.4pt}1111\discretionary{/}{%
}{/}eufm\hspace{.1pt}\discretionary{.}{%
}{.}\hspace{.4pt}12067}}}


\bibitem{he2024videopro}
J.~He, X.~Wang, W.~Kam-Kwai, X.~Huang, C.~Chen, Z.~Chen, F.~Wang, M.~Zhu, and H.~Qu.
\newblock {VideoPro: A Visual Analytics Approach for Interactive Video Programming}.
\newblock {\em IEEE Transactions on Visualization and Computer Graphics}, 30(1):87--97, 2024. \href{https://doi.org/10.1109/TVCG.2023.3326586}
{doi: {{%
10\hspace{.1pt}\discretionary{.}{%
}{.}\hspace{.4pt}1109\discretionary{/}{%
}{/}TVCG\hspace{.1pt}\discretionary{.}{%
}{.}\hspace{.4pt}2023\hspace{.1pt}\discretionary{.}{%
}{.}\hspace{.4pt}3326586}}}


\bibitem{related_stock_app_hullman_2013}
J.~Hullman, N.~Diakopoulos, and E.~Adar.
\newblock {Contextifier: Automatic Generation of Annotated Stock Visualizations}.
\newblock In {\em Proc. CHI}, pp. 2707--2716. ACM, New York, 2013. \href{https://doi.org/10.1145/2470654.2481374}
{doi: {{%
10\hspace{.1pt}\discretionary{.}{%
}{.}\hspace{.4pt}1145\discretionary{/}{%
}{/}2470654\hspace{.1pt}\discretionary{.}{%
}{.}\hspace{.4pt}2481374}}}


\bibitem{related_stock_market_joseph_2013}
J.~Joseph and I.~Indratmo.
\newblock {Visualizing Stock Market Data with Self-Organizing Map}.
\newblock In {\em Proc. FLAIRS}, 2013.

\bibitem{wong2023anchorage}
W.~Kam-Kwai, X.~Wang, Y.~Wang, J.~He, R.~Zhang, and H.~Qu.
\newblock {Anchorage: Visual Analysis of Satisfaction in Customer Service Videos Via Anchor Events}.
\newblock {\em IEEE Transactions on Visualization and Computer Graphics}, 30(7):4008--4022, 2024. \href{https://doi.org/10.1109/TVCG.2023.3245609}
{doi: {{%
10\hspace{.1pt}\discretionary{.}{%
}{.}\hspace{.4pt}1109\discretionary{/}{%
}{/}TVCG\hspace{.1pt}\discretionary{.}{%
}{.}\hspace{.4pt}2023\hspace{.1pt}\discretionary{.}{%
}{.}\hspace{.4pt}3245609}}}


\bibitem{related_stock_pairwise_keim_2006}
D.~A. Keim, T.~Nietzschmann, N.~Schelwies, J.~Schneidewind, T.~Schreck, and H.~Ziegler.
\newblock {A Spectral Visualization System for Analyzing Financial Time Series Data}.
\newblock In {\em Proc. EUROVIS}. The Eurographics Association, 2006. \href{https://doi.org/10.2312/VisSym/EuroVis06/195-202}
{doi: {{%
10\hspace{.1pt}\discretionary{.}{%
}{.}\hspace{.4pt}2312\discretionary{/}{%
}{/}VisSym\discretionary{/}{%
}{/}EuroVis06\discretionary{/}{%
}{/}195\discretionary{%
}{-}{-}202}}}


\bibitem{related_clustering_kern_2017}
M.~Kern, A.~Lex, N.~Gehlenborg, and C.~R. Johnson.
\newblock {Interactive Visual Exploration and Refinement of Cluster Assignments}.
\newblock {\em BMC bioinformatics}, 18(1):1--13, 2017. \href{https://doi.org/10.1186/s12859-017-1813-7}
{doi: {{%
10\hspace{.1pt}\discretionary{.}{%
}{.}\hspace{.4pt}1186\discretionary{/}{%
}{/}s12859\discretionary{%
}{-}{-}017\discretionary{%
}{-}{-}1813\discretionary{%
}{-}{-}7}}}


\bibitem{related_stock_survey_ko_2016}
S.~Ko, I.~Cho, S.~Afzal, C.~Yau, J.~Chae, A.~Malik, K.~Beck, Y.~Jang, W.~Ribarsky, and D.~S. Ebert.
\newblock {A Survey on Visual Analysis Approaches for Financial Data}.
\newblock {\em Computer Graphics Forum}, 35(3):599--617, 2016. \href{https://doi.org/10.1111/cgf.12931}
{doi: {{%
10\hspace{.1pt}\discretionary{.}{%
}{.}\hspace{.4pt}1111\discretionary{/}{%
}{/}cgf\hspace{.1pt}\discretionary{.}{%
}{.}\hspace{.4pt}12931}}}


\bibitem{related_clustering_kown_2018}
B.~C. {Kwon}, B.~{Eysenbach}, J.~{Verma}, K.~{Ng}, C.~{De Filippi}, W.~F. {Stewart}, and A.~{Perer}.
\newblock {Clustervision: Visual Supervision of Unsupervised Clustering}.
\newblock {\em IEEE Transactions on Visualization and Computer Graphics}, 24(1):142--151, 2018. \href{https://doi.org/10.1109/TVCG.2017.2745085}
{doi: {{%
10\hspace{.1pt}\discretionary{.}{%
}{.}\hspace{.4pt}1109\discretionary{/}{%
}{/}TVCG\hspace{.1pt}\discretionary{.}{%
}{.}\hspace{.4pt}2017\hspace{.1pt}\discretionary{.}{%
}{.}\hspace{.4pt}2745085}}}


\bibitem{related_stock_market_lei_2010}
S.~T. Lei and K.~Zhang.
\newblock {A Visual Analytics System for Financial Time-Series Data}.
\newblock In {\em Proc. VINCI}. ACM, New York, 2010. \href{https://doi.org/10.1145/1865841.1865868}
{doi: {{%
10\hspace{.1pt}\discretionary{.}{%
}{.}\hspace{.4pt}1145\discretionary{/}{%
}{/}1865841\hspace{.1pt}\discretionary{.}{%
}{.}\hspace{.4pt}1865868}}}


\bibitem{related_network_leite_2020}
R.~A. Leite, T.~Gschwandtner, S.~Miksch, E.~Gstrein, and J.~Kuntner.
\newblock {Neva: Visual Analytics to Identify Fraudulent Networks}.
\newblock {\em Computer Graphics Forum}, 39(6):344--359, 2020. \href{https://doi.org/10.1111/cgf.14042}
{doi: {{%
10\hspace{.1pt}\discretionary{.}{%
}{.}\hspace{.4pt}1111\discretionary{/}{%
}{/}cgf\hspace{.1pt}\discretionary{.}{%
}{.}\hspace{.4pt}14042}}}


\bibitem{design_matrix_lex_2014}
A.~{Lex}, N.~{Gehlenborg}, H.~{Strobelt}, R.~{Vuillemot}, and H.~{Pfister}.
\newblock {UpSet: Visualization of Intersecting Sets}.
\newblock {\em IEEE Transactions on Visualization and Computer Graphics}, 20(12):1983--1992, 2014. \href{https://doi.org/10.1109/TVCG.2014.2346248}
{doi: {{%
10\hspace{.1pt}\discretionary{.}{%
}{.}\hspace{.4pt}1109\discretionary{/}{%
}{/}TVCG\hspace{.1pt}\discretionary{.}{%
}{.}\hspace{.4pt}2014\hspace{.1pt}\discretionary{.}{%
}{.}\hspace{.4pt}2346248}}}


\bibitem{related_network_lin_2021}
Y.~{Lin}, K.~{Wong}, Y.~{Wang}, R.~{Zhang}, B.~{Dong}, H.~{Qu}, and Q.~{Zheng}.
\newblock {TaxThemis: Interactive Mining and Exploration of Suspicious Tax Evasion Groups}.
\newblock {\em IEEE Transactions on Visualization and Computer Graphics}, 27(2):849--859, 2021. \href{https://doi.org/10.1109/TVCG.2020.3030370}
{doi: {{%
10\hspace{.1pt}\discretionary{.}{%
}{.}\hspace{.4pt}1109\discretionary{/}{%
}{/}TVCG\hspace{.1pt}\discretionary{.}{%
}{.}\hspace{.4pt}2020\hspace{.1pt}\discretionary{.}{%
}{.}\hspace{.4pt}3030370}}}


\bibitem{background_cluster_liu_2020}
Y.~Liu, L.~He, D.~Li, X.~Luo, G.~Peng, X.~Fan, and G.~Sun.
\newblock {Correlation Analysis of Chinese Pork Concept Stocks Based on Big Data}.
\newblock In {\em Artificial Intelligence and Security}, pp. 475--486. Springer, Cham, 2020. \href{https://doi.org/10.1007/978-3-030-57881-7_42}
{doi: {{%
10\hspace{.1pt}\discretionary{.}{%
}{.}\hspace{.4pt}1007\discretionary{/}{%
}{/}978\discretionary{%
}{-}{-}3\discretionary{%
}{-}{-}030\discretionary{%
}{-}{-}57881\discretionary{%
}{-}{-}7\_42}}}


\bibitem{design_network_loretan_2000}
M.~Loretan and W.~B. English.
\newblock {Evaluating Correlation Breakdowns during Periods of Market Volatility}.
\newblock {\em Available at SSRN 231857}, 2000.

\bibitem{introduction_mantegna_1999}
R.~N. Mantegna.
\newblock {Hierarchical Structure in Financial Markets}.
\newblock {\em The European Physical Journal B-Condensed Matter and Complex Systems}, 11(1):193--197, 1999. \href{https://doi.org/10.1007/s100510050929}
{doi: {{%
10\hspace{.1pt}\discretionary{.}{%
}{.}\hspace{.4pt}1007\discretionary{/}{%
}{/}s100510050929}}}


\bibitem{task_marti_2021}
G.~Marti, F.~Nielsen, M.~Bi{\'{n}}kowski, and P.~Donnat.
\newblock {\em {A Review of Two Decades of Correlations, Hierarchies, Networks and Clustering in Financial Markets}}, pp. 245--274.
\newblock Springer, Cham, 2021. \href{https://doi.org/10.1007/978-3-030-65459-7_10}
{doi: {{%
10\hspace{.1pt}\discretionary{.}{%
}{.}\hspace{.4pt}1007\discretionary{/}{%
}{/}978\discretionary{%
}{-}{-}3\discretionary{%
}{-}{-}030\discretionary{%
}{-}{-}65459\discretionary{%
}{-}{-}7\_10}}}


\bibitem{system_network_mcgee_2019}
F.~McGee, M.~Ghoniem, G.~Melançon, B.~Otjacques, and B.~Pinaud.
\newblock {The State of the Art in Multilayer Network Visualization}.
\newblock {\em Computer Graphics Forum}, 38(6):125--149, 2019. \href{https://doi.org/10.1111/cgf.13610}
{doi: {{%
10\hspace{.1pt}\discretionary{.}{%
}{.}\hspace{.4pt}1111\discretionary{/}{%
}{/}cgf\hspace{.1pt}\discretionary{.}{%
}{.}\hspace{.4pt}13610}}}


\bibitem{task_munzner_2009}
T.~Munzner.
\newblock {A Nested Model for Visualization Design and Validation}.
\newblock {\em IEEE Transactions on Visualization and Computer Graphics}, 15(6):921--928, 2009. \href{https://doi.org/10.1109/TVCG.2009.111}
{doi: {{%
10\hspace{.1pt}\discretionary{.}{%
}{.}\hspace{.4pt}1109\discretionary{/}{%
}{/}TVCG\hspace{.1pt}\discretionary{.}{%
}{.}\hspace{.4pt}2009\hspace{.1pt}\discretionary{.}{%
}{.}\hspace{.4pt}111}}}


\bibitem{related_knowledge_nam_2009}
J.~E. Nam, M.~Maurer, and K.~Mueller.
\newblock {A High-Dimensional Feature Clustering Approach to Support Knowledge-Assisted Visualization}.
\newblock {\em Computers \& Graphics}, 33(5):607--615, 2009. \href{https://doi.org/10.1016/j.cag.2009.06.006}
{doi: {{%
10\hspace{.1pt}\discretionary{.}{%
}{.}\hspace{.4pt}1016\discretionary{/}{%
}{/}j\hspace{.1pt}\discretionary{.}{%
}{.}\hspace{.4pt}cag\hspace{.1pt}\discretionary{.}{%
}{.}\hspace{.4pt}2009\hspace{.1pt}\discretionary{.}{%
}{.}\hspace{.4pt}06\hspace{.1pt}\discretionary{.}{%
}{.}\hspace{.4pt}006}}}


\bibitem{related_network_niu_2020}
Z.~{Niu}, R.~{Li}, J.~{Wu}, D.~{Cheng}, and J.~{Zhang}.
\newblock {iConViz: Interactive Visual Exploration of the Default Contagion Risk of Networked-Guarantee Loans}.
\newblock In {\em Proc. VAST}, pp. 84--94. IEEE, 2020. \href{https://doi.org/10.1109/VAST50239.2020.00013}
{doi: {{%
10\hspace{.1pt}\discretionary{.}{%
}{.}\hspace{.4pt}1109\discretionary{/}{%
}{/}VAST50239\hspace{.1pt}\discretionary{.}{%
}{.}\hspace{.4pt}2020\hspace{.1pt}\discretionary{.}{%
}{.}\hspace{.4pt}00013}}}


\bibitem{introduction_onnela_2003}
J.-P. Onnela, A.~Chakraborti, K.~Kaski, J.~Kert\'esz, and A.~Kanto.
\newblock {Dynamics of Market Correlations: Taxonomy and Portfolio Analysis}.
\newblock {\em Phys. Rev. E}, 68:056110, Nov. 2003. \href{https://doi.org/10.1103/PhysRevE.68.056110}
{doi: {{%
10\hspace{.1pt}\discretionary{.}{%
}{.}\hspace{.4pt}1103\discretionary{/}{%
}{/}PhysRevE\hspace{.1pt}\discretionary{.}{%
}{.}\hspace{.4pt}68\hspace{.1pt}\discretionary{.}{%
}{.}\hspace{.4pt}056110}}}


\bibitem{related_clustering_pister_2021}
A.~{Pister}, P.~{Buono}, J.~D. {Fekete}, C.~{Plaisant}, and P.~{Valdivia}.
\newblock {Integrating Prior Knowledge in Mixed-Initiative Social Network Clustering}.
\newblock {\em IEEE Transactions on Visualization and Computer Graphics}, 27(2):1775--1785, 2021. \href{https://doi.org/10.1109/TVCG.2020.3030347}
{doi: {{%
10\hspace{.1pt}\discretionary{.}{%
}{.}\hspace{.4pt}1109\discretionary{/}{%
}{/}TVCG\hspace{.1pt}\discretionary{.}{%
}{.}\hspace{.4pt}2020\hspace{.1pt}\discretionary{.}{%
}{.}\hspace{.4pt}3030347}}}


\bibitem{task_podobnik_2009}
B.~Podobnik, D.~Horvatic, A.~M. Petersen, and H.~E. Stanley.
\newblock {Cross-Correlations between Volume Change and Price Change}.
\newblock {\em PNAS}, 106(52):22079--22084, 2009. \href{https://doi.org/10.1073/pnas.0911983106}
{doi: {{%
10\hspace{.1pt}\discretionary{.}{%
}{.}\hspace{.4pt}1073\discretionary{/}{%
}{/}pnas\hspace{.1pt}\discretionary{.}{%
}{.}\hspace{.4pt}0911983106}}}


\bibitem{related_stock_market_rea_2014}
A.~Rea and W.~Rea.
\newblock {Visualization of a Stock Market Correlation Matrix}.
\newblock {\em Physica A: Statistical Mechanics and its Applications}, 400:109--123, 2014. \href{https://doi.org/10.1016/j.physa.2014.01.017}
{doi: {{%
10\hspace{.1pt}\discretionary{.}{%
}{.}\hspace{.4pt}1016\discretionary{/}{%
}{/}j\hspace{.1pt}\discretionary{.}{%
}{.}\hspace{.4pt}physa\hspace{.1pt}\discretionary{.}{%
}{.}\hspace{.4pt}2014\hspace{.1pt}\discretionary{.}{%
}{.}\hspace{.4pt}01\hspace{.1pt}\discretionary{.}{%
}{.}\hspace{.4pt}017}}}


\bibitem{related_clustering_sacha_2018}
D.~{Sacha}, M.~{Kraus}, J.~{Bernard}, M.~{Behrisch}, T.~{Schreck}, Y.~{Asano}, and D.~A. {Keim}.
\newblock {SOMFlow: Guided Exploratory Cluster Analysis with Self-Organizing Maps and Analytic Provenance}.
\newblock {\em IEEE Transactions on Visualization and Computer Graphics}, 24(1):120--130, 2018. \href{https://doi.org/10.1109/TVCG.2017.2744805}
{doi: {{%
10\hspace{.1pt}\discretionary{.}{%
}{.}\hspace{.4pt}1109\discretionary{/}{%
}{/}TVCG\hspace{.1pt}\discretionary{.}{%
}{.}\hspace{.4pt}2017\hspace{.1pt}\discretionary{.}{%
}{.}\hspace{.4pt}2744805}}}


\bibitem{system_data_shen_2017}
Q.~{Shen}, T.~{Wu}, H.~{Yang}, Y.~{Wu}, H.~{Qu}, and W.~{Cui}.
\newblock {NameClarifier: A Visual Analytics System for Author Name Disambiguation}.
\newblock {\em IEEE Transactions on Visualization and Computer Graphics}, 23(1):141--150, 2017. \href{https://doi.org/10.1109/TVCG.2016.2598465}
{doi: {{%
10\hspace{.1pt}\discretionary{.}{%
}{.}\hspace{.4pt}1109\discretionary{/}{%
}{/}TVCG\hspace{.1pt}\discretionary{.}{%
}{.}\hspace{.4pt}2016\hspace{.1pt}\discretionary{.}{%
}{.}\hspace{.4pt}2598465}}}


\bibitem{related_stock_app_shi_2019}
L.~{Shi}, Z.~{Teng}, L.~{Wang}, Y.~{Zhang}, and A.~{Binder}.
\newblock {DeepClue: Visual Interpretation of Text-Based Deep Stock Prediction}.
\newblock {\em IEEE Transactions on Knowledge and Data Engineering}, 31(6):1094--1108, 2019. \href{https://doi.org/10.1109/TKDE.2018.2854193}
{doi: {{%
10\hspace{.1pt}\discretionary{.}{%
}{.}\hspace{.4pt}1109\discretionary{/}{%
}{/}TKDE\hspace{.1pt}\discretionary{.}{%
}{.}\hspace{.4pt}2018\hspace{.1pt}\discretionary{.}{%
}{.}\hspace{.4pt}2854193}}}


\bibitem{related_stock_pairwise_simon_2018}
P.~M. Simon and C.~Turkay.
\newblock {Hunting High and Low: Visualising Shifting Correlations in Financial Markets}.
\newblock {\em Computer Graphics Forum}, 37(3):479--490, 2018. \href{https://doi.org/10.1111/cgf.13435}
{doi: {{%
10\hspace{.1pt}\discretionary{.}{%
}{.}\hspace{.4pt}1111\discretionary{/}{%
}{/}cgf\hspace{.1pt}\discretionary{.}{%
}{.}\hspace{.4pt}13435}}}


\bibitem{design_prism_sips_2012}
M.~{Sips}, P.~{Köthur}, A.~{Unger}, H.~{Hege}, and D.~{Dransch}.
\newblock {A Visual Analytics Approach to Multiscale Exploration of Environmental Time Series}.
\newblock {\em IEEE Transactions on Visualization and Computer Graphics}, 18(12):2899--2907, 2012. \href{https://doi.org/10.1109/TVCG.2012.191}
{doi: {{%
10\hspace{.1pt}\discretionary{.}{%
}{.}\hspace{.4pt}1109\discretionary{/}{%
}{/}TVCG\hspace{.1pt}\discretionary{.}{%
}{.}\hspace{.4pt}2012\hspace{.1pt}\discretionary{.}{%
}{.}\hspace{.4pt}191}}}


\bibitem{related_stock_pairwise_stitz_2016}
H.~{Stitz}, S.~{Gratzl}, W.~{Aigner}, and M.~{Streit}.
\newblock {ThermalPlot: Visualizing Multi-Attribute Time-Series Data Using a Thermal Metaphor}.
\newblock {\em IEEE Transactions on Visualization and Computer Graphics}, 22(12):2594--2607, 2016. \href{https://doi.org/10.1109/TVCG.2015.2513389}
{doi: {{%
10\hspace{.1pt}\discretionary{.}{%
}{.}\hspace{.4pt}1109\discretionary{/}{%
}{/}TVCG\hspace{.1pt}\discretionary{.}{%
}{.}\hspace{.4pt}2015\hspace{.1pt}\discretionary{.}{%
}{.}\hspace{.4pt}2513389}}}


\bibitem{introduction_tola_2008}
V.~Tola, F.~Lillo, M.~Gallegati, and R.~N. Mantegna.
\newblock {Cluster Analysis for Portfolio Optimization}.
\newblock {\em Journal of Economic Dynamics and Control}, 32(1):235--258, 2008. \href{https://doi.org/10.1016/j.jedc.2007.01.034}
{doi: {{%
10\hspace{.1pt}\discretionary{.}{%
}{.}\hspace{.4pt}1016\discretionary{/}{%
}{/}j\hspace{.1pt}\discretionary{.}{%
}{.}\hspace{.4pt}jedc\hspace{.1pt}\discretionary{.}{%
}{.}\hspace{.4pt}2007\hspace{.1pt}\discretionary{.}{%
}{.}\hspace{.4pt}01\hspace{.1pt}\discretionary{.}{%
}{.}\hspace{.4pt}034}}}


\bibitem{system_correlation_tumminello_2010}
M.~Tumminello, F.~Lillo, and R.~N. Mantegna.
\newblock {Correlation, Hierarchies, and Networks in Financial Markets}.
\newblock {\em Journal of Economic Behavior \& Organization}, 75(1):40--58, 2010. \href{https://doi.org/10.1016/j.jebo.2010.01.004}
{doi: {{%
10\hspace{.1pt}\discretionary{.}{%
}{.}\hspace{.4pt}1016\discretionary{/}{%
}{/}j\hspace{.1pt}\discretionary{.}{%
}{.}\hspace{.4pt}jebo\hspace{.1pt}\discretionary{.}{%
}{.}\hspace{.4pt}2010\hspace{.1pt}\discretionary{.}{%
}{.}\hspace{.4pt}01\hspace{.1pt}\discretionary{.}{%
}{.}\hspace{.4pt}004}}}


\bibitem{related_clustering_turkay_2011}
C.~Turkay, J.~Parulek, N.~Reuter, and H.~Hauser.
\newblock {Interactive Visual Analysis of Temporal Cluster Structures}.
\newblock {\em Computer Graphics Forum}, 30(3):711--720, 2011. \href{https://doi.org/10.1111/j.1467-8659.2011.01920.x}
{doi: {{%
10\hspace{.1pt}\discretionary{.}{%
}{.}\hspace{.4pt}1111\discretionary{/}{%
}{/}j\hspace{.1pt}\discretionary{.}{%
}{.}\hspace{.4pt}1467\discretionary{%
}{-}{-}8659\hspace{.1pt}\discretionary{.}{%
}{.}\hspace{.4pt}2011\hspace{.1pt}\discretionary{.}{%
}{.}\hspace{.4pt}01920\hspace{.1pt}\discretionary{.}{%
}{.}\hspace{.4pt}x}}}


\bibitem{system_cluster_wang_2020}
H.~Wang, Y.~Yang, and B.~Liu.
\newblock {GMC: Graph-Based Multi-View Clustering}.
\newblock {\em IEEE Transactions on Knowledge and Data Engineering}, 32(6):1116--1129, 2020. \href{https://doi.org/10.1109/TKDE.2019.2903810}
{doi: {{%
10\hspace{.1pt}\discretionary{.}{%
}{.}\hspace{.4pt}1109\discretionary{/}{%
}{/}TKDE\hspace{.1pt}\discretionary{.}{%
}{.}\hspace{.4pt}2019\hspace{.1pt}\discretionary{.}{%
}{.}\hspace{.4pt}2903810}}}


\bibitem{related_stock_market_wattenberg_1999}
M.~Wattenberg.
\newblock {Visualizing the Stock Market}.
\newblock In {\em Proc. CHI EA}, pp. 188--189. ACM, New York, 1999. \href{https://doi.org/10.1145/632716.632834}
{doi: {{%
10\hspace{.1pt}\discretionary{.}{%
}{.}\hspace{.4pt}1145\discretionary{/}{%
}{/}632716\hspace{.1pt}\discretionary{.}{%
}{.}\hspace{.4pt}632834}}}


\bibitem{related_clustering_xia_2023}
J.~Xia, L.~Huang, W.~Lin, X.~Zhao, J.~Wu, Y.~Chen, Y.~Zhao, and W.~Chen.
\newblock {Interactive Visual Cluster Analysis by Contrastive Dimensionality Reduction}.
\newblock {\em IEEE Transactions on Visualization and Computer Graphics}, 29(1):734--744, 2023. \href{https://doi.org/10.1109/TVCG.2022.3209423}
{doi: {{%
10\hspace{.1pt}\discretionary{.}{%
}{.}\hspace{.4pt}1109\discretionary{/}{%
}{/}TVCG\hspace{.1pt}\discretionary{.}{%
}{.}\hspace{.4pt}2022\hspace{.1pt}\discretionary{.}{%
}{.}\hspace{.4pt}3209423}}}


\bibitem{yang2023examples}
L.~Yang, C.~Xiong, W.~Kam-Kwai, A.~Wu, and H.~Qu.
\newblock {Explaining with Examples Lessons Learned from Crowdsourced Introductory Description of Information Visualizations}.
\newblock {\em IEEE Transactions on Visualization and Computer Graphics}, 29(3):1638--1650, 2023. \href{https://doi.org/10.1109/TVCG.2021.3128157}
{doi: {{%
10\hspace{.1pt}\discretionary{.}{%
}{.}\hspace{.4pt}1109\discretionary{/}{%
}{/}TVCG\hspace{.1pt}\discretionary{.}{%
}{.}\hspace{.4pt}2021\hspace{.1pt}\discretionary{.}{%
}{.}\hspace{.4pt}3128157}}}


\bibitem{yuan2023taxscheduler}
L.~Yuan, B.~Li, S.~Li, W.~Kam-Kwai, R.~Zhang, and H.~Qu.
\newblock {Tax-Scheduler: An Interactive Visualization System for Staff Shifting and Scheduling at Tax Authorities}.
\newblock {\em Visual Informatics}, 7(2):30--40, June 2023. \href{https://doi.org/10.1016/j.visinf.2023.02.001}
{doi: {{%
10\hspace{.1pt}\discretionary{.}{%
}{.}\hspace{.4pt}1016\discretionary{/}{%
}{/}j\hspace{.1pt}\discretionary{.}{%
}{.}\hspace{.4pt}visinf\hspace{.1pt}\discretionary{.}{%
}{.}\hspace{.4pt}2023\hspace{.1pt}\discretionary{.}{%
}{.}\hspace{.4pt}02\hspace{.1pt}\discretionary{.}{%
}{.}\hspace{.4pt}001}}}


\bibitem{related_stock_app_yue_2020}
X.~{Yue}, J.~{Bai}, Q.~{Liu}, Y.~{Tang}, A.~{Puri}, K.~{Li}, and H.~{Qu}.
\newblock {sPortfolio: Stratified Visual Analysis of Stock Portfolios}.
\newblock {\em IEEE Transactions on Visualization and Computer Graphics}, 26(1):601--610, 2020. \href{https://doi.org/10.1109/TVCG.2019.2934660}
{doi: {{%
10\hspace{.1pt}\discretionary{.}{%
}{.}\hspace{.4pt}1109\discretionary{/}{%
}{/}TVCG\hspace{.1pt}\discretionary{.}{%
}{.}\hspace{.4pt}2019\hspace{.1pt}\discretionary{.}{%
}{.}\hspace{.4pt}2934660}}}


\bibitem{related_network_yue_2019}
X.~{Yue}, X.~{Shu}, X.~{Zhu}, X.~{Du}, Z.~{Yu}, D.~{Papadopoulos}, and S.~{Liu}.
\newblock {BitExTract: Interactive Visualization for Extracting Bitcoin Exchange Intelligence}.
\newblock {\em IEEE Transactions on Visualization and Computer Graphics}, 25(1):162--171, 2019. \href{https://doi.org/10.1109/TVCG.2018.2864814}
{doi: {{%
10\hspace{.1pt}\discretionary{.}{%
}{.}\hspace{.4pt}1109\discretionary{/}{%
}{/}TVCG\hspace{.1pt}\discretionary{.}{%
}{.}\hspace{.4pt}2018\hspace{.1pt}\discretionary{.}{%
}{.}\hspace{.4pt}2864814}}}


\bibitem{zhang2024scrolltimes}
W.~Zhang, W.~Kam-Kwai, Y.~Chen, A.~Jia, L.~Wang, J.-W. Zhang, L.~Cheng, H.~Qu, and W.~Chen.
\newblock {ScrollTimes: Tracing the Provenance of Paintings as a Window Into History}.
\newblock {\em IEEE Transactions on Visualization and Computer Graphics}, 30(6):2981--2994, 2024. \href{https://doi.org/10.1109/TVCG.2024.3388523}
{doi: {{%
10\hspace{.1pt}\discretionary{.}{%
}{.}\hspace{.4pt}1109\discretionary{/}{%
}{/}TVCG\hspace{.1pt}\discretionary{.}{%
}{.}\hspace{.4pt}2024\hspace{.1pt}\discretionary{.}{%
}{.}\hspace{.4pt}3388523}}}


\bibitem{zhang2023cohortva}
W.~Zhang, W.~Kam-Kwai, X.~Wang, Y.~Gong, R.~Zhu, K.~Liu, Z.~Yan, S.~Tan, H.~Qu, S.~Chen, and W.~Chen.
\newblock {CohortVA: A Visual Analytic System for Interactive Exploration of Cohorts based on Historical Data}.
\newblock {\em IEEE Transactions on Visualization and Computer Graphics}, 29(1):756--766, 2023. \href{https://doi.org/10.1109/TVCG.2022.3209483}
{doi: {{%
10\hspace{.1pt}\discretionary{.}{%
}{.}\hspace{.4pt}1109\discretionary{/}{%
}{/}TVCG\hspace{.1pt}\discretionary{.}{%
}{.}\hspace{.4pt}2022\hspace{.1pt}\discretionary{.}{%
}{.}\hspace{.4pt}3209483}}}


\bibitem{related_stock_market_ziegler_2010}
H.~{Ziegler}, M.~{Jenny}, T.~{Gruse}, and D.~A. {Keim}.
\newblock {Visual Market Sector Analysis for Financial Time Series Data}.
\newblock In {\em Proc. VAST}, pp. 83--90, 2010. \href{https://doi.org/10.1109/VAST.2010.5652530}
{doi: {{%
10\hspace{.1pt}\discretionary{.}{%
}{.}\hspace{.4pt}1109\discretionary{/}{%
}{/}VAST\hspace{.1pt}\discretionary{.}{%
}{.}\hspace{.4pt}2010\hspace{.1pt}\discretionary{.}{%
}{.}\hspace{.4pt}5652530}}}


\end{thebibliography}
